\documentclass[floatfix,reprint,aps,prc,showpacs,showkeys,eqsecnum,amsmath,amsfonts,amssymb,twocolumn,groupedaddress,superscriptaddress,nofootinbib,noeprint]{revtex4-2}

\usepackage{graphicx}    % Include figure files
\usepackage{float} 

\usepackage{dcolumn}     % Align table columns on decimal point
\usepackage{bm}          % bold math
\usepackage[hyperindex,breaklinks]{hyperref}

\usepackage{amsmath} 
\usepackage{amssymb} 
\usepackage{amsfonts} 
\usepackage{latexsym}
\usepackage{enumitem} % Used to reduce itemize/enumerate spacing

\usepackage{booktabs} % Top and bottom rules for tables

\usepackage{cleveref}

\usepackage[english]{babel}
\usepackage[expansion=all,babel]{microtype}

\usepackage[usenames]{xcolor}

\usepackage{mathtools}

\usepackage{multirow}

\usepackage{rotating}

\newcommand{\sat}[1]{{#1}_{\mathrm{sat}}}
\newcommand{\sym}[1]{{#1}_{\mathrm{sym}}}
\newcommand{\asym}[1]{{#1}_{\mathrm{asym}}}
\newcommand{\eff}[1]{{#1}_{\mathrm{eff}}}

\newcommand{\mn}{m_{\mathrm{eff;\,n}}^{\mathrm {PNM}}}
\newcommand{\mni}[1]{m_{\mathrm{eff;\,n;}\,#1}^{\mathrm{PNM}}}

\newcommand{\mN}{m_{\mathrm{eff;\,N}}^{\mathrm {SNM}}}
\newcommand{\mNi}[1]{m_{\mathrm{eff;\,N;}\,#1}^{\mathrm{SNM}}}

\usepackage{txfonts}
\newcommand{\be}{\begin{equation}}
\newcommand{\ee}{\end{equation}}
\newcommand{\ba}{\begin{eqnarray}}
\newcommand{\ea}{\end{eqnarray}}
\newcommand{\tth}[1]{{#1}_{\mathrm{th}}}

\newcommand{\Msun}{M_\odot}

\makeatletter
	\newcommand{\vast}{\bBigg@{2.85}}
\makeatother

\begin{document}
%%%%%%%%%%%%%%%%%%%%%%%%%%%%%%%%%%%%%%%%%%%%%%%%%%%%%%%%%%%%%%%%%%%

\title{Bayesian inference of the dense matter equation of state built upon extended Skyrme interactions: A generalization}

\preprint{Grant\#}

\author{Mikhail V. Beznogov}
\email{mikhail.beznogov@nipne.ro}
\affiliation{National Institute for Physics and Nuclear Engineering (IFIN-HH), RO-077125 Bucharest, Romania}

\author{Adriana R. Raduta}
\email{araduta@nipne.ro}
\affiliation{National Institute for Physics and Nuclear Engineering (IFIN-HH), RO-077125 Bucharest, Romania}

\date{\today}% It is always \today, today,

%%%%%%%%%%%%%%%%%%%%%%%%%%%%%%%%%%%%%%%%%%%%%%%%%%%%%%%%%%%%%%%%%%%
%%%%%%%%%%%%%%%%%%%%%%%%%%%%%%%%%%%%%%%%%%%%%%%%%%%%%%%%%%%%%%%%%%%
\begin{abstract}
  The nonrelativistic theory of nuclear matter (NM) based on Brussels-Skyrme interactions is employed to develop models for dense and neutron-rich matter within a Bayesian framework.
  We employ the following set of constraints: the four best-known nuclear empirical parameters, the density dependence of the energy per particle in pure neutron matter (PNM), the density dependence of the Landau effective mass ($\eff{m}$) of neutrons in PNM and symmetric NM, and a lower limit on the maximum gravitational mass that neutron stars (NSs) can support.
  In addition, a number of ``sanity checks'' are added: the values of the speed of sound, neutron and proton Landau effective masses, and Fermi velocities are constrained up to the central density of the most massive NS configuration and for isospin asymmetries $\delta=(n_n-n_p)/(n_n+n_p)$ ranging from 0 to 1.
  Our ensemble of models \emph{fully} explores the capacity of nonrelativistic  Brussels-Skyrme effective interactions to describe NM at densities exceeding several times the nuclear saturation density.
  This is a necessary step toward a better understanding of the properties of dense matter and possible correlations between the parameters of NSs and the parameters of NM.
  Due to the pronounced U-shaped density dependencies of $\eff{m}$, which qualitatively agree with the predictions of ab initio calculations featuring three-body forces, all our models exhibit a non-monotonic ``rise-and-fall'' behavior of the thermal pressure ($\tth{p}$) as a function of density, which in extreme cases leads to $\tth{p}<0$.
  This work is a generalization of [Beznogov and Raduta, Phys. Rev. C 110, 035805 (2024)].
\end{abstract}

\maketitle
%%%%%%%%%%%%%%%%%%%%%%%%%%%%%%%%%%%%%%%%%%%%%%%%%%%%%%%%%%%%%%%%%%%

%%%%%%%%%%%%%%%%%%%%%%%%%%%%%%%%%%%%%%%%%%%%%%%%%%%%%%%%%%%%%%%%%%%%%%
\section{Introduction}
\label{sec:Intro}
%%%%%%%%%%%%%%%%%%%%%%%%%%%%%%%%%%%%%%%%%%%%%%%%%%%%%%%%%%%%%%%%%%%%%%

The last decade has seen unprecedented progress in the multi-messenger astronomy of neutron stars (NSs).
Precise measurements of massive pulsars' masses~\cite{Antoniadis_Science_2013, Cromartie_Nature_2020}, estimates of NSs' radii via X-ray pulse profile modeling~\cite{Miller_ApJ_2019,Riley_ApJL_2019,Miller_ApJL_2021,Riley_ApJL_2021,Salmi_ApJ_2022,Salmi_ApJ_2024b,Vinciguerra_ApJ_2024,Miller_ApJ_2026}, and the inference of (combined) tidal deformability via gravitational wave (GW) detection~\cite{Abbott_PRL_2017,Abbott_PRL_121,Abbott_PRX_2019} have opened new avenues for understanding matter at extreme densities and isospin asymmetries that are inaccessible in terrestrial laboratories.

Thus, a large variety of phenomenological models, traditionally employed to address properties of finite nuclei and nuclear matter (NM), e.g., nonrelativistic mean-field models with zero-range interactions~\cite{Papakonstantinou_PRC_2023,Imam_PRD_2024,Beznogov_ApJ_2024,Beznogov_PRC_2024} and relativistic mean-field models with non-linear and/or density-dependent couplings~\cite{Traversi_ApJ_2020,Malik_ApJ_2022,Beznogov_PRC_2023,Malik_PRD_2023,Char_PRD_2023,Imam_PRD_2024,Li_PRC_2025,Cartaxo_ApJSS_2026}, have been used to study the behavior of dense and neutron-rich NM, compatibility with available astrophysics constraints and possible drawbacks caused by the use of energy density functionals (EDFs) of limited flexibility.
All these recent studies account for constraints from nuclear physics experiments, most often translated into values taken by the nuclear empirical parameters (NEPs), ab initio calculations of the energy per particle and/or pressure in pure neutron matter (PNM) at densities up to the saturation density of symmetric matter ($\sat{n} \approx 0.16~\mathrm{fm}^{-3} \approx 2 \cdot 10^{14}~\mathrm{g/cm^3}$), and a lower boundary on the maximum gravitational mass of NSs of $\approx 2~\Msun$.
In some cases, constraints from measurements of NSs' radii and tidal deformabilities, or results of perturbative quantum chromodynamics, valid for $n \gtrsim 40 \sat{n}$, were also included. 
The heterogeneity of these constraints, combined with their dissimilar and poorly understood sensitivities to various density domains and isospin asymmetries, as well as the convoluted shapes of credibility regions (CRs), makes Bayesian techniques the method of choice.
The ensemble of these works allowed researchers to understand the roles of the underlying effective interaction, priors, the set of constraints, the likelihood, and Markov Chain Monte Carlo (MCMC) sampling details, as well as to distinguish between physical and spurious correlations among properties of NM and/or global parameters of NSs. 

The aim of this paper is to extend and generalize the work done in Refs.~\cite{Beznogov_ApJ_2024,Beznogov_PRC_2024}, hereafter referred to as Paper~I and Paper~II, respectively.
Paper~I focused on the Bayesian inference of the equation of state (EoS) of dense matter using EoS models built within the nonrelativistic mean-field model of NM with standard Skyrme interactions~\cite{Negele_PRC_1972}. 
It employed a minimal set of constraints consisting of the four best-known NEPs, the density dependence of the energy per neutron ($E/A$) in PNM, and a lower limit on the maximum mass of NSs, in addition to physicality conditions on NS matter (i.e., thermodynamical stability and causality).
In some runs, we also constrained the neutron effective masses ($m_{\mathrm{eff};n}$) in symmetric nuclear matter (SNM) and PNM at $n=0.16~\mathrm{fm}^{-3}$, while in others, we accounted for correlations between the values of $E/A$ in PNM  at different densities.
A significant number of the EoS models thus generated feature nucleon effective masses that, for densities in excess of a few times $\sat{n}$, reach values as low as $\sim 0.1 m_n$ (where $m_n$ stands for the bare mass of the neutron) and/or symmetry energies ($\sym{E}$) that have rise-and-fall behavior as a function of density ($n$).
A significant drawback of these low $m_{\mathrm{eff};n}$ values is that the Fermi velocity $v_{F;n}$ can exceed the speed of light ($c$)~\cite{Urban_PRC_2023}, thereby rendering the model unphysical. 
Low values of $\sym{E}$ are incompatible with the direct Urca processes operating in nucleonic stars~\cite{Fortin_PRC_2016}.
Moreover, the rise-and-fall behavior of $\sym{E}(n)$ is in tension with the predictions of relativistic mean-field models that utilize the same constraints~\cite{Malik_ApJ_2022,Beznogov_PRC_2023,Malik_PRD_2023,Li_PRC_2025,Cartaxo_ApJSS_2026}.

Paper~II repeated the analysis of Paper~I but employed a Brussels extended Skyrme interaction~\cite{Chamel_PRC_2009}.
Additional momentum-dependent terms allow for a more complex density dependence of $m_{\mathrm{eff}}$ and provide greater flexibility for the EDF.
This prevents $m_{\mathrm{eff}}$ from decreasing excessively, which is a necessary condition for $v_{F;n}/c<1$,  and can even produce a $U$-shaped density dependence that qualitatively agrees with the predictions of Brueckner-Hartree-Fock~\cite{Baldo_PRC_2014,Burgio_PRC_2020} and $\chi$EFT~\cite{Drischler_PRC_2021} calculations featuring three-body forces.
The density dependence of $m_{\mathrm{eff}}$ is of particular interest for numerical simulations of core-collapse supernovae and binary NS mergers because it determines the thermal response of matter~\cite{Constantinou_PRC_2014,Constantinou_PRC_2015}.
Models allowing $m_{\mathrm{eff}}$ to increase with density are noteworthy because they accommodate negative thermal pressure over wide domains of temperature, density, and proton fraction~\cite{Raduta_AA_2026}.
The increased flexibility of the EDF allows NEPs and global parameters of NSs to span broader domains and erases spurious correlations among NEPs or between NEPs and parameters of NSs.
Paper~II showed that in most cases $\sym{E}$ increases with density.
Nevertheless, Paper~II did not fully exploit the flexibility of the Brussels extended Skyrme interactions, nor did it guarantee that NS matter and PNM remain physical up to the central density of the most massive NS configuration ($n_c^*$).
These two limitations resulted from a trade-off to increase the computational efficiency of the Bayesian inference.
Specifically, in Paper~II we set $x_4=x_5=0$, where $x_4$ and $x_5$ are parameters entering the extra momentum terms, and  replaced $n_c^*$ by a lower value $n_l=0.8~\mathrm{fm^{-3}}$.

In this paper, we address these limitations.
The models developed here explore the full range of behaviors allowed by the Brussels extended Skyrme interaction; they are compatible with the constraints and remain physical up to $n_c^*=1.0 - 1.2~\mathrm{fm}^{-3}$, including for PNM.

The remainder of the paper has the following structure. Sec.~\ref{sec:Model} offers a brief review of the model. Details on Bayesian inference are given in Sec.~\ref{sec:Setup}. Sec.~\ref{sec:Results} discusses the results.
Conclusions are drawn in Sec.~\ref{sec:Concl}.

%%%%%%%%%%%%%%%%%%%%%%%%%%%%%%%%%%%%%%%%%%%%%%%%%%%%%%%%%%%%%%%%%%%%%%
\section{The model}
\label{sec:Model}
%%%%%%%%%%%%%%%%%%%%%%%%%%%%%%%%%%%%%%%%%%%%%%%%%%%%%%%%%%%%%%%%%%%%%%

As in Paper~II, we employ the Brussels extended Skyrme interactions.
They are obtained by supplementing the standard Skyrme interactions~\cite{Negele_PRC_1972}
\begin{align}
	V\left({\mathbf r}_1, {\mathbf r}_2 \right) &= t_0 \left(1+x_0 P_{\sigma} \right) \delta\left( {\mathbf r}\right) \nonumber \\
	&+ \frac{t_1}2  \left(1+x_1 P_{\sigma} \right) \left[{\mathbf k}'^2 \delta\left( {\mathbf r}\right) +\delta\left( {\mathbf r}\right) {\mathbf k}^2 \right] \nonumber \\
    &+ t_2 \left(1+x_2 P_{\sigma} \right) {\mathbf k}' \cdot \delta\left( {\mathbf r}\right) {\mathbf k}
    \label{eq:VSk}\\
	&+ \frac{t_3}6 \left(1+x_3 P_{\sigma} \right) \left[n \left({\mathbf R} \right)\right]^{\sigma} \delta\left( {\mathbf r}\right) \nonumber \\
    &+ i W_0 \left( \boldsymbol{\sigma}_1 + \boldsymbol{\sigma}_2\right) \cdot \left[ {\mathbf k}' \times \delta\left({\mathbf r}\right) {\mathbf k} \right], \nonumber
\end{align}
with two density-dependent terms~\cite{Chamel_PRC_2009},
\begin{align}
    & V^{\mathrm{BSk}}\left({\mathbf r}_1, {\mathbf r}_2 \right) = V\left({\mathbf r}_1, {\mathbf r}_2 \right) \nonumber \\
    &+ \frac{1}2 \left(t_4+t_4 x_4 P_{\sigma} \right) \left[{\mathbf k}'^2 \left[n \left({\mathbf R} \right)\right]^{\beta} \delta\left( {\mathbf r}\right) + \delta\left( {\mathbf r}\right)  \left[n \left({\mathbf R} \right)\right]^{\beta} {\mathbf k}^2 \right]
    \label{eq:VBSk} \\
    &+\left(t_5+t_5 x_5 P_{\sigma} \right){\mathbf k}'\cdot \left[n \left({\mathbf R} \right)\right]^{\gamma} \delta\left( {\mathbf r}\right) {\mathbf k}~. \nonumber
\end{align}

In Eqs.~\eqref{eq:VSk} and \eqref{eq:VBSk}, ${\mathbf r}_i$ denotes the coordinates of the interacting nucleons ($i=1,2$) ${\mathbf r}={\mathbf r}_1-{\mathbf r}_2$ and ${\mathbf R}=\left({\mathbf r}_1+{\mathbf r}_2\right)/2$; ${\mathbf k}=\left(\boldsymbol{\nabla}_1- \boldsymbol{\nabla}_2\right)/2i$ is the relative momentum operator acting on the right and ${\mathbf k}'$ is its conjugate acting on the left; $P_{\sigma}=\left( 1+ \boldsymbol{\sigma}_1 \cdot \boldsymbol{\sigma}_2\right)/2$ is the two-body spin-exchange operator; $n({\mathbf r})=n_\mathrm{n}({\mathbf r})+n_\mathrm{p}({\mathbf r})$ is the total local density; $n_j({\mathbf r})$, with $j=\mathrm{n},\mathrm{p}$, are the neutron and proton local densities.
$t_0$, $t_1$, $t_2$, $t_3$, $t_4$, $t_5$, $x_0$, $x_1$, $x_2$, $x_3$, $t_4 x_4$, $t_5 x_5$, $\sigma$, $\beta$, $\gamma$, and $W_0$ are parameters determined from fits to experimental and theoretical data.
In the absence of spin polarization, the parameter $W_0$ is irrelevant and is therefore omitted hereafter.

A considerable number of such effective interactions have been proposed and successfully used in theoretical nuclear physics for the description of both finite nuclei and NM.
They are particularly appealing for studies of cold NM because analytic expressions in terms of total density ($n=n_n+n_p$), isovector density ($n_3=n_n-n_p$), and parameters of the interaction can be obtained for most of the quantities of interest.
These include the energy density ($e$), pressure ($P$), Landau effective masses ($m_{\mathrm{eff};i}$) of neutrons ($i=n$) and protons ($i=p$), neutron and proton Fermi velocities ($v_{F;i}=\hbar k_{F;i}/m_{\mathrm{eff};i}$), as well as the NEPs.
$\hbar k_{F;i}$ is the Fermi momentum of $i$-nucleons, and $k_{F;i}=(3 \pi^2 n_i)^{1/3}$.

NEPs collectively refer to the saturation density of SNM ($\sat{n}$) and the coefficients in the Taylor expansion of the energy per particle ($E/A=e/n$) in terms of the departure $x=(n-\sat{n})/3 \sat{n}$ from $\sat{n}$ and the isospin asymmetry $\delta=n_3/n=1-2Y_p$, where $Y_p=n_p/n$ represents the proton fraction,
\be
E(n,\delta)/A=E_0(n,0)+\sum_{i=2,4,...} \delta^{i} E_{\mathrm{sym};i}(n,0),
\ee
with
\ba
E_0(n,0)=\sum_{j=0,1,2,...} \frac1{j!} X_{\mathrm{sat}}^{j} x^j,
\label{eq:isos}\\
E_{\mathrm{sym};i}(n,0)=\sum_{k=0,1,2,...}\frac{1}{k!} X_{\mathrm{sym};i}^{k} x^k, ~~~ i=2,4,...,
\label{eq:isov}
\ea
where
\ba
\sat{X}^{j}&=&3^j \sat{n}^j \left. \left( \frac{\partial^j (e/n)}{\partial n^j}\right) \right|_{n=\sat{n}}, \\
X_{\mathrm{sym};i}^{k}&=&\left.\left(\frac{\partial^k E_{\mathrm{sym};i}(n,0)}{\partial x^k} \right)\right|_{n=n_{\mathrm{sat}}}.
\ea

The lowest-order terms in Eqs.~\eqref{eq:isos} and \eqref{eq:isov} computed at $\sat{n}$ represent the saturation energy of SNM, $\sat{E} \equiv E_0(\sat{n},0)$, and the symmetry energy
\be
\sym{J}=E_{\mathrm{sym};2}(\sat{n})=\frac12 \left.\frac{\partial^2 (e/n)}{\partial \delta^2}\right|_{\sat{n}, \delta=0},
\ee
respectively. 
The first-, second-, third-, and fourth-order coefficients in the Taylor expansions in Eqs.~\eqref{eq:isos} and \eqref{eq:isov} are customarily denoted in the literature by $L$, $K$, $Q$ and $Z$ with superscripts that indicate whether they refer to expansions in the isoscalar or isovector sectors.
Together with $\sym{J}$, $\sym{L}$, ..., $\sym{Z}$, the behavior of asymmetric NM is gauged by the asymmetry energy, defined as the per-nucleon cost of converting SNM into PNM, $\asym{E}(n)=E(n,\delta=1)/A-E(n,\delta=0)/A$.  

To complete the survey of cold NM with the Brussels extended Skyrme interactions, we recall that the extra momentum-dependent terms in Eq.~\eqref{eq:VBSk} allow the Landau effective mass
\ba
    \frac{1}{m_{\mathrm{eff};i}} &=& \frac{1}{k_i} \left. \frac{d e_i}{d k_i}\right|_{k_i=k_{F;\,i}}
    \label{eq:meff:def}\\
    &=& \frac{1}{m_i} + \frac{2}{\hbar^2} \left( \widetilde{C}_{\mathrm{eff}}(n) n \pm \widetilde{D}_{\mathrm{eff}} (n) n_3 \right),
\label{eq:meff:BSk}
\ea
to have a complex density dependence. 
In Eq.~\eqref{eq:meff:BSk}, the ``+'' and ``$-$''  signs correspond to $i=n$ and $i=p$, respectively, and
\begin{align}
	\begin{split}
		\widetilde{C}_{\mathrm{eff}}(n) &= \frac{1}{16}\Big[3 t_1 + t_2 \left(4 x_2 + 5 \right)\Big] \\
		&+ \frac{1}{16} \Big[3 t_4 n^\beta + \left(4 t_5 x_5 +5 t_5\right) n^\gamma \Big],
	\end{split}
	\label{eq:Ceff} \\
	\begin{split}
		\widetilde{D}_{\mathrm{eff}} (n) &= \frac{1}{16}\Big[t_2 \left(2 x_2 + 1\right) - t_1 \left(2 x_1 + 1\right) \Big]  \\
		&+ \frac{1}{16}\Big[-\left(2 t_4 x_4 + t_4\right) n^\beta + \left(2 t_5 x_5 + t_5\right) n^\gamma \Big].
	\end{split}
	\label{eq:Deff}
\end{align}
The $U$-shaped density dependencies obtained for a wide range of parameters are of particular interest, as they qualitatively agree with the predictions of Brueckner-Hartree-Fock calculations with three-body forces~\cite{Baldo_PRC_2014,Burgio_PRC_2020} and $\chi$EFT models~\cite{Drischler_PRC_2021}. 
In addition, the issue of superluminal Fermi velocities of nucleons $(v_{F;i} >c)$, which plagues most standard Skyrme interactions~\cite{Urban_PRC_2023}, can be easily resolved.

Alternative extensions of the Skyrme interaction exist as well.
One of the most intensively considered extensions was proposed by Agrawal et al.~\cite{Agrawal_PRC_2006} and consists of replacing the term $t_3 \left(1+x_3 P_{\sigma} \right) \left[n \left({\mathbf R} \right)\right]^{\sigma} \delta\left( {\mathbf r}\right)/6$ in Eq.~\eqref{eq:VSk} with a sum $\sum_{i=1}^3 t_{3i} \left(1+x_{3i} P_{\sigma} \right) \left[n \left({\mathbf R} \right)\right]^{\sigma_i} \delta\left( {\mathbf r}\right)/6$.
The presence of several density-dependent terms with different powers of the density obviously increases the flexibility of the EDF.
However, as none of these terms enters the expression for the Landau effective mass, see Eq.~\eqref{eq:meff:BSk}, the parametrization in Ref.~\cite{Agrawal_PRC_2006} suffers from the same drawbacks as the standard one.

In both standard and extended Skyrme, $m_{\mathrm{eff};i}$ is independent of temperature. Expressions for thermal energy,
\begin{align}
    \begin{split}
        \tth{e}(n, \delta, T) &= e(n, \delta, T)-e(n, \delta, T=0) \\
        &=\sum_{i=n,p} \frac{\hbar^2}{2 m_{\mathrm{eff};i}} [ \tau_i (T)-\tau_i(T=0) ],
    \end{split}
    \label{eq:eth}
\end{align}
and thermal pressure,
\begin{align}
    \begin{split}
    &\tth{P}(n, \delta, T) = P(n, \delta, T)-P(n, \delta, T=0) \\
    &= \sum_{i=n,p} \frac{\hbar^2}{3 m_{\mathrm{eff};i}}
    \left(1-\frac32 \frac{n}{m_{\mathrm{eff};i}} \frac{\partial m_{\mathrm{eff};i}}{\partial n}\right)
    [ \tau_i (T)-\tau_i(T=0) ],
    \end{split}
    \label{eq:pth}    
\end{align}
demonstrate that the thermal response is governed by $m_{\mathrm{eff};i}(n)$.
In Eqs.~\eqref{eq:eth} and \eqref{eq:pth}, $\tau_i$ stands for the density of kinetic energy.
In the low-temperature limit, this dependence can be worked out using the Sommerfeld expansion.
Considering that at the lowest order in temperature,
\be
\tau_i(T)-\tau_i(T=0)=\frac{T^2}{\hbar^4} \left(\frac{\pi}{3} \right)^{2/3}  m_{\mathrm{eff};i}^2 n_i^{1/3},
\ee
Eqs.~\eqref{eq:eth} and \eqref{eq:pth} become~\cite{Raduta_AA_2026}
\ba
\tth{e} &\approx& \frac{T^2}{2 \hbar^2}  \left(\frac{\pi}{3} \right)^{2/3} \sum_{i=n,p} m_{\mathrm{eff};i} n_i^{1/3},\\
\tth{P} &\approx& \frac{T^2}{3 \hbar^2}  \left(\frac{\pi}{3} \right)^{2/3} \sum_{i=n,p} \left[ m_{\mathrm{eff};i} -\frac32 \frac{\partial m_{\mathrm{eff};i} }{\partial n}n \right] n_i^{1/3}. 
\ea

For a compilation of expressions at zero temperature, see Paper~I, Paper~II, and Ref.~\cite{Dutra_PRC_2012}. 
For a review of thermal effects in models with zero-range Skyrme interactions, see Ref.~\cite{Constantinou_PRC_2014,Constantinou_PRC_2015}.

%%%%%%%%%%%%%%%%%%%%%%%%%%%%%%%%%%%%%%%%%%%%%%%%%%%%%%%%%%%%%%%%%%%
\section{The Bayesian Setup}
\label{sec:Setup}
%%%%%%%%%%%%%%%%%%%%%%%%%%%%%%%%%%%%%%%%%%%%%%%%%%%%%%%%%%%%%%%%%%%%%%

%%%%%%%%%%%%%%%%%%%%%%%%%%%%%%%%%%%%%%%%%%%%%%%%%%%%%%%%%%%%%%%%%%%%%%
\subsection{Constraints}
\label{ssec:Constraints}
%%%%%%%%%%%%%%%%%%%%%%%%%%%%%%%%%%%%%%%%%%%%%%%%%%%%%%%%%%%%%%%%%%%%%%

%%%%%%%%%%%%%%%%%%%%%%%%%%%%%%%%%%%%%%%%%%%%%%%%%%%%%%%%%%%%%%%%%%%%%%
\renewcommand{\arraystretch}{1.20}
\setlength{\tabcolsep}{10pt}
\begin{table}
    \centering
    \caption{Constraints imposed on the EOS models.
    $\sat{E}$ and $\sat{K}$ represent the energy per particle and compression modulus of symmetric saturated matter at density $\sat{n}$; 
    $\sym{J}$ stands for the symmetry energy at saturation;
    $\left(E/A\right)_i$ with $i=2,3,4$ stand for the energy per particle of PNM at the densities of 0.08, 0.12, and 0.16 $\mathrm{fm}^{-3}$;
    $\mni{i}$ ($\mNi{i}$) denotes the Landau effective mass of the neutron (nucleon) in PNM (SNM) at density $n_i$;
    $M_{\mathrm{G}}^*$ represents the maximum gravitational mass of an NS;
    $M_{\mathrm{DU}}$ is the lower limit of an NS gravitational mass that accommodates the direct Urca process.
    For all quantities except $M_{\mathrm{G}}^*$ and $M_{\mathrm{DU}}$, we provide the mean and the corresponding standard deviation;
    for $M_{\mathrm{G}}^*$ and $M_{\mathrm{DU}}$, we specify the threshold values.
    $m_\mathrm{n}$ and $m_\mathrm{N}$ represent the neutron and nucleon bare masses, respectively.
    Note that, with the exception of $M_{\mathrm{DU}}$, the same values were used in Paper~II.}
    \begin{tabular*}{\columnwidth}{ccccc}
        \toprule
        \toprule
        Quantity             & Units              & Value   & Std. deviation & Ref. \\
        \midrule
        $n_{\mathrm{sat}}$   & $\mathrm{fm}^{-3}$ & 0.16 	& 0.004         & a \\
        $E_{\mathrm{sat}}$   & MeV 		          & $-15.9$ & 0.2 	        & a \\
        $K_{\mathrm{sat}}$   & MeV 		          & 240 	& 30 	        & a \\ 
        $J_{\mathrm{sym}}$   & MeV 		          & 30.8   & 1.6            & a  \\
        $\left(E/A\right)_2$ & MeV 	              & 9.212  &  0.226         & b \\
        $\left(E/A\right)_3$ & MeV 	              & 12.356 &  0.512         & b \\
        $\left(E/A\right)_4$ & MeV 	              & 15.877 &  0.872         & b \\       
        $\mNi{2}$            & $m_\mathrm{N}$     & 0.715  &  0.011         & b \\       
        $\mNi{3}$            & $m_\mathrm{N}$     & 0.667  &  0.012         & b \\       
        $\mNi{4}$            & $m_\mathrm{N}$     & 0.638  &  0.013         & b \\       
        $\mni{2}$            & $m_\mathrm{n}$     & 0.877  &  0.004         & b \\
        $\mni{3}$            & $m_\mathrm{n}$     & 0.866  &  0.011         & b \\
        $\mni{4}$            & $m_\mathrm{n}$     & 0.880  &  0.026         & b \\
        $M_{\mathrm{G}}^*$   & $\Msun$ 	          & $>2.0$  &  --          & c  \\
        $M_{\mathrm{DU}}$   & $\Msun$ 	          & $>1.5$  &  --          & --  \\
        \bottomrule
        \bottomrule
    \end{tabular*}
    \label{tab:constraints}
    {\raggedright \textbf{References.} (a) \citet{Margueron_PRC_2018a}; (b) \citet{Drischler_PRC_2021}; (c) \citet{Fonseca_2021}.\par}
\end{table}
\setlength{\tabcolsep}{2.0pt}
\renewcommand{\arraystretch}{1.0}
%%%%%%%%%%%%%%%%%%%%%%%%%%%%%%%%%%%%%%%%%%%%%%%%%%%%%%%%%%%%%%%%%%%%%%

Following the approach of Paper~I, Paper~II, and Ref.~\cite{Beznogov_PRC_2023}, we employ a minimal set of constraints derived from nuclear physics data, $\chi$EFT calculations of SNM and PNM with densities up to $\sat{n}$, and astrophysical observations. 
To ensure that all EoS models are physical, a series of ``sanity checks'' was also implemented.

The nuclear physics constraints correspond to the saturation density ($\sat{n}$), the energy per particle($\sat{E}$), and the compression modulus ($\sat{K}$) of saturated SNM, as well as the symmetry energy ($\sym{J}$) at $\sat{n}$, which are the best-known NEPs. 
As in Paper~II, we use the values calculated by Margueron et~al.~\cite{Margueron_PRC_2018a} based on a set of 35 frequently used Skyrme interactions. 
For specific values, see Table~\ref{tab:constraints}. 

The behavior of PNM and SNM over $0.08 \leq n\,[\mathrm{fm}^{-3}] \leq 0.16$ is constrained by the $\chi$EFT calculations in Ref.~\cite{Drischler_PRC_2021}, where nucleon-nucleon (NN) interactions computed at N$^3$LO are supplemented by three-nucleon (3N) interactions computed at N$^2$LO.
For PNM (SNM), we employ constraints on the energy per particle ($E/A$) and neutron (nucleon) effective mass ($\mn$ ($\mN$)) for three values of density. 
For details, see Table~\ref{tab:constraints}.

As in previous works, we require NS EoSs to provide maximum gravitational masses in excess of $2~\Msun$.
We further require that the electron direct Urca process, which comprises the $\beta$-decay of neutrons ($n \to p+e^-+\bar\nu_e$) and the electron capture on protons ($p+e^- \to n+\nu_e$), occurs in NSs with masses of at least $1.5~\Msun$.
This condition is imposed to comply with astrophysical data on NSs' thermal evolution, which seem to require (at least one) fast-cooling process~\cite{Beznogov_MNRAS_2015a,Beznogov_MNRAS_2015b} in sufficiently heavy NSs, but not in canonical-mass (i.e., $1.4~\Msun$) NSs. 

Extra constraints implemented to ensure physical consistency are as follows:
(C1) for $n \leq n_c^*$, NS matter and PNM are causal ($c_S^2 \leq c^2$).
Here and throughout the paper, $n_c^*$ represents the central density of the maximum mass configuration and $c_S^2$ stands for the speed of sound squared;
(C2) for $n \leq n_c^*$, neutron and proton effective masses in PNM and SNM take values between 0 and the corresponding bare masses;
(C3) for $n \leq n_c^*$, the Fermi velocity of neutrons ($v_{F;n}$) in PNM and SNM does not exceed $c$;
(C4) NS matter is thermodynamically stable, i.e., $P>0$ and $dP/dn \geq 0$, where $P$ represents pressure.
Since SNM is softer than PNM, the causality of SNM (condition (C1)) does not need to be imposed explicitly.
Similarly, the thermodynamic stability of NS matter automatically ensures the thermodynamic stability of NM with arbitrary isospin asymmetry; we have explicitly verified this fact. 

%%%%%%%%%%%%%%%%%%%%%%%%%%%%%%%%%%%%%%%%%%%%%%%%%%%%%%%%%%%%%%%%%%%%%%
\subsection{Likelihood}
\label{ssec:Chi}
%%%%%%%%%%%%%%%%%%%%%%%%%%%%%%%%%%%%%%%%%%%%%%%%%%%%%%%%%%%%%%%%%%%%%%

The likelihood is implemented as in Paper~II, but assuming no correlations between constraints.
Consequently, there are only two types of constraints: uncorrelated probabilistic constraints and threshold (hard-wall) constraints. 
NEPs and energy per particle in PNM fall into the first category; maximum NS mass, the direct Urca threshold NS mass, and ``sanity checks'' fall into the second.
Effective mass constraints are treated as probabilistic for densities up to $\sat{n}$ and as threshold constraints for higher densities.

The log-likelihood function is the sum of these two parts:
\begin{align}
\log \mathcal{L}=\log \mathcal{L}_{\mathrm{uncorr.}} + \log \mathcal{L}_{\mathrm{thresh.}}.
	\label{eq:LogL}
\end{align}
For the uncorrelated constraints, the expression is
\begin{align}
	\log \mathcal{L}_{\mathrm{uncorr.}} \propto -\chi_{\mathrm{uncorr.}}^2 = -\frac{1}{2} \sum_{i=1}^{N_\mathrm{uncorr.}}  \left(\frac{d_i - \xi_i(\mathbf{\Theta}) } {\mathcal{Z}_i} \right)^2,
	\label{eq:Chi2-NoCorr}
\end{align}
where $N_\mathrm{uncorr.}$ is the number of uncorrelated constraints; $d_i$ and $\mathcal{Z}_i$ stand for the constraint and its standard deviation,respectively; and $\xi_i(\mathbf{\Theta})$ is the value predicted by the model (defined by the parameter set $\mathbf{\Theta}$) for quantity $i$. 
For the meaning and values of $d_i$ and $\mathcal{Z}_i$, see Table~\ref{tab:constraints}.

The threshold constraints are implemented as:
\begin{align}
    \log \mathcal{L}_{\mathrm{thresh.}} \propto -\chi_{\mathrm{thresh.}}^2 = 
    \begin{cases}
        0,         & \text{condition satisfied} \\
        -10^{10},  &\text{condition violated} 
     \end{cases}
\label{eq:Chi2-Thr}
\end{align}
%

%%%%%%%%%%%%%%%%%%%%%%%%%%%%%%%%%%%%%%%%%%%%%%%%%%%%%%%%%%%%%%%%%%%%%%
\subsection{Priors}
\label{ssec:Priors}
%%%%%%%%%%%%%%%%%%%%%%%%%%%%%%%%%%%%%%%%%%%%%%%%%%%%%%%%%%%%%%%%%%%%%%

%%%%%%%%%%%%%%%%%%%%%%%%%%%%%%%%%%%%%%%%%%%%%%%%%%%%%%%%%%%%%%%%%%%%%%
\renewcommand{\arraystretch}{1.05}
\setlength{\tabcolsep}{8.0pt}
\begin{table}
	%\vspace{0.4cm}
	\caption{Domains of the priors.}
	\centering
	\begin{tabular}{cccc}
		\toprule
		\toprule
		Parameter      & Units                          & Min.              & Max.         \\
		\midrule
		$\sat{n}$      & $\mathrm{fm^{-3}}$             & 0.14              & 0.18          \\
		$\sat{E}$      & MeV                           & $-16.9$           & $-14.9$       \\
		$\sym{J}$      & MeV                           & 22.8              & 38.8          \\
		$D_3$          & MeV $\mathrm{fm^{3+3\sigma}}$   & $-1\,500$         & 500           \\
		$\eff{C}$      & MeV $\mathrm{fm^{5}}$          & $-500$            & 2\,500        \\
		$\eff{D}$      & MeV $\mathrm{fm^{5}}$          & $-1\,000$         & 2\,000        \\
		$t_4$	       & MeV $\mathrm{fm^{5+3\beta}}$	& $-500$			& 500  \\
		$t_5$	       & MeV $\mathrm{fm^{5+3\gamma}}$    & $-500$		    & 500      \\
		$t_4 x_4$      & MeV $\mathrm{fm^{5+3\beta}}$	& $-5\,000$			& 5\,000		   \\
		$t_5 x_5$      & MeV $\mathrm{fm^{5+3\gamma}}$    & $-5\,000$		    & 5\,000		   \\
		$\beta$	       &  ---				& 0.07		   & 1.1   \\
		$\gamma$       &  ---				& 0.07		   & 1.1   \\
		$\sigma$       &  ---                           & 0.07             & 1.1   \\
		\bottomrule
		\bottomrule
	\end{tabular}
	\label{tab:Prior}
\end{table}
\setlength{\tabcolsep}{2.0pt}
\renewcommand{\arraystretch}{1.0}
%%%%%%%%%%%%%%%%%%%%%%%%%%%%%%%%%%%%%%%%%%%%%%%%%%%%%%%%%%%%%%%%%%%%%%

As in Paper~II, a ``mixed'' input parametrization is used, i.e., instead of sampling all 13 parameters of the effective interaction directly, we sample three NEPs ($\sat{n}$, $\sat{E}$, and $\sym{J}$) and 10 interaction parameters ($D_3$, $\eff{C}$, $\eff{D}$, $t_4$, $t_5$, $t_4 x_4$, $t_5 x_5$, $\sigma$, $\beta$, $\gamma$). 
This coordinate transformation significantly simplifies the sampling with minimal impact on the posterior distributions; see Appendix~B of Paper~I.
The remaining parameters of the effective interaction ($C_0$, $D_0$, and $C_3$) are calculated according to Eqs.~(3.6) of Paper~II.

As discussed in Paper~II, the parameters $t_4$ and $t_4 x_4$ (as well as $t_5$ and $t_5 x_5$) always enter all expressions together except for $\sym{J}$. 
This poses a problem for sampling because, if their priors are wide enough, they can always cancel each other out. 
We observed this effect in our test runs: as we widened the priors for $t_4$, $t_4 x_4$, $t_5$, and $t_5 x_5$, their posteriors became self-similar and broadened with the priors.
Thus, we had to limit the width of the prior for these four parameters somewhat arbitrarily. 
From a series of BSk models (BSk18 -- BSk26~\cite{Chamel_PRC_2009,BSk19-BSk21,BSk22-BSk26} and BSk28 -- BSk32~\cite{BSk29-BSk32}), it follows that $t_4$ and $t_5$ vary roughly from $-500$ to $-40$, $x_4$ varies from $-6$ to $5$, 
and $x_5$ varies from $-13$ to $-2$. 
We have chosen the prior ranges for these four parameters accordingly: from $-500$ to $500$ for $t_4$ and $t_5$, and from $-5000$ to $5000$ for $t_4 x_4$ and $t_5 x_5$.
We have also widened the prior for $D_3$ compared to Paper~II.
Other priors remained unchanged.
Note also that, compared to Paper~II, we made a slight additional transformation from $\{t_4, x_4, t_5, x_5\}$ to $\{t_4, t_4 x_4, t_5, t_5 x_5\}$. 
This transformation is trivial.

The domains of the priors are provided in Table~\ref{tab:Prior}, and we sample uniformly from them. 

%%%%%%%%%%%%%%%%%%%%%%%%%%%%%%%%%%%%%%%%%%%%%%%%%%%%%%%%%%%%%%%%%%%%%%
\subsection{Sampling}
\label{ssec:Sampling}
%%%%%%%%%%%%%%%%%%%%%%%%%%%%%%%%%%%%%%%%%%%%%%%%%%%%%%%%%%%%%%%%%%%%%%

The general setup of our two-stage inference procedure is exactly the same as in Papers~I and II. 
We refer the interested reader to Appendix~B of Paper~I for more details.
We employed the \textsc{ptemcee} sampler (v.1.0.0)~\cite{emcee,ptemcee}\,\footnote{The documentation is available at \url{https://emcee.readthedocs.io/en/v2.2.1/user/pt/} and the GitHub page is \url{https://github.com/willvousden/ptemcee}. Note that \textsc{ptemcee} is no longer maintained.} with tempered chains as our main inference tool.
The configuration included 7 tempered chains with 1\,000 walkers each, a maximum temperature of $T=16384$, and automatic temperature adjustments for intermediate chains.
The latter were necessary to equalize the inter-chain swap probabilities, which stabilized at approximately $33\%$ after the initial burn-in phase.
We made sure that even the highest temperature chain could not violate hard-wall constraints.
As discussed in Appendix~B of Paper~II, with tempered chains, the autocorrelation length was not an issue, as the inter-chain swaps erase intra-chain correlations.
Similarly to Paper~II, we thinned the chains until the fraction of repetitions in the main chain fell well below $10^{-4}$.

We verified the stability of the posterior distributions using bootstrap analysis and additionally cross-checked them by independently computing them via the MLFriends~\cite{Buchner_2016,Buchner_2019} nested sampling~\cite{Skilling_2004,Skilling_2006} algorithm, implemented in the Python package \textsc{ultranest} (v.4.5.0)~\cite{ultranest}\,\footnote{The documentation is available at \url{https://johannesbuchner.github.io/UltraNest/} and the GitHub page is \url{https://github.com/JohannesBuchner/UltraNest}.}. 
The \textsc{ultranest} code was custom-modified in the same manner as described in Appendix~B of Paper~II.
The posteriors obtained via \textsc{ptemcee} and \textsc{ultranest} were compared and found to be very similar, confirming the robustness of the results.

%%%%%%%%%%%%%%%%%%%%%%%%%%%%%%%%%%%%%%%%%%%%%%%%%%%%%%%%%%%%%%%%%%%%%%
\section{Results}
\label{sec:Results}
%%%%%%%%%%%%%%%%%%%%%%%%%%%%%%%%%%%%%%%%%%%%%%%%%%%%%%%%%%%%%%%%%%%%%%

The following subsections discuss the properties of NM in terms of $\mn$, $\mN$, energy per particle ($E/A$) in SNM and PNM, and $\asym{E}$ as functions of density, alongside the marginalized posterior distributions of NEPs, $E/A$ in PNM, $\mn$, and $\mN$. 
NS properties are discussed in terms of the density dependence of the speed of sound squared and the proton fraction, as well as marginalized posterior distributions of key global observables.
We also evaluate compliance with astrophysical inferences on the radii of $1.4~\Msun$ and $2~\Msun$ NSs and the tidal deformability of $1.4~\Msun$ NSs, which were not included in the likelihood.
Special attention is given to the correlations between NEPs and NS parameters, as well as  to comparisons with existing literature.

\looseness=-1
The EoS models generated in this work are systematically compared with the ensemble and a subclass of EoS models from run~2 of Paper~II, hereafter referred to as ``11\,NF'' and ``11\,F,'' respectively. 
``11'' refers to the number of non-fixed parameters of the effective interaction in Paper~II; ``F'' and ``NF'' stand for ``filtered'' and ``not-filtered''. 
The EoS models in 11\,NF were obtained as described in Sec.~\ref{ssec:Constraints}, except that:
no constraint was imposed on $M_{\mathrm{DU}}$;
for (C1), the causality of PNM was disregarded;
the condition (C2) was imposed only up to a density $n_l=0.8~\mathrm{fm}^{-3}$;
the condition (C3) was imposed only for neutrons and only up to $n_l$.
The EoS models in 11\,F comply with all the constraints in Sec.~\ref{ssec:Constraints}, except for the one on $M_{\mathrm{DU}}$.

The models built in this work explore the ability of the Brussels extended Skyrme interaction to describe the behavior of matter with densities up to $\sim 7.5\sat{n}$, which is well beyond the original target.
A comparison with the models in 11\,F allows one to gauge the role of the terms involving $t_4 x_4$ and $t_5 x_5$, which were suppressed in Paper~II, as well as the role of the $M_{\mathrm{DU}}$ constraint.
A comparison between the models in 11\,F and those in 11\,NF probes the consequences of extending the validity domain of the model from $n_l \approx 5\sat{n}$ to $n_c^* \approx 7.5 \sat{n}$, and from matter in $\beta$-equilibrium to PNM.

%%%%%%%%%%%%%%%%%%%%%%%%%%%%%%%%%%%%%%%%%%%%%%%%%%%%%%%%%%%%%%%%%%%%%%
\subsection{Nuclear matter}
\label{ssec:NM}
%%%%%%%%%%%%%%%%%%%%%%%%%%%%%%%%%%%%%%%%%%%%%%%%%%%%%%%%%%%%%%%%%%%%%%

%%%%%%%%%%%%%%%%%%%%%%%%%%%%%%%%%%%%%%%%%%%%%%%%%%%%%%%%%%%%%%%%%%%%%%
\begin{figure}
  \centering
  \includegraphics[]{"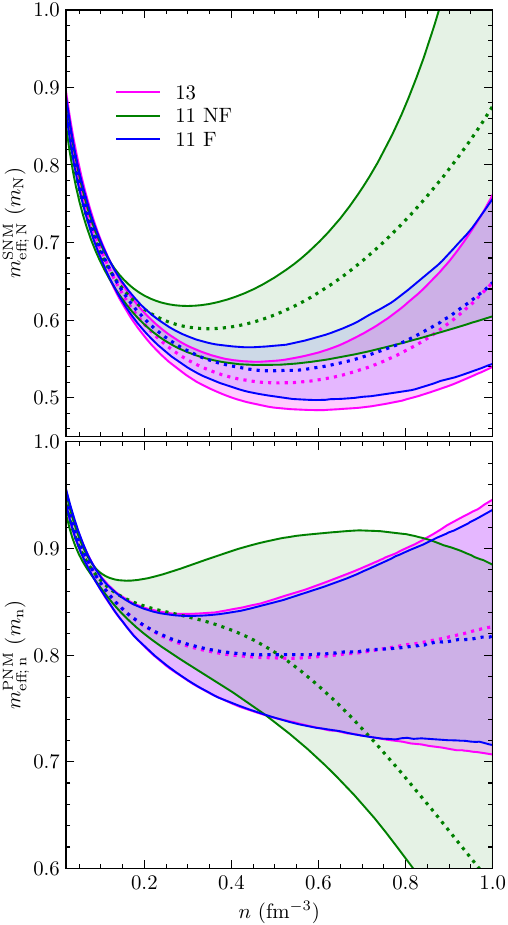"}
  \caption{
  Nucleon effective mass ($\mN$) in SNM (top) and neutron effective mass ($\mn$) in PNM (bottom) as functions of density.
  Medians (dotted curves) and the bounds of the 90\% CI (solid lines) for three sets of models.
  For details, see Sec.~\ref{sec:Results}.
  }
  \label{Fig:CD_meff_NM}
\end{figure}
%%%%%%%%%%%%%%%%%%%%%%%%%%%%%%%%%%%%%%%%%%%%%%%%%%%%%%%%%%%%%%%%%%%%%%

%%%%%%%%%%%%%%%%%%%%%%%%%%%%%%%%%%%%%%%%%%%%%%%%%%%%%%%%%%%%%%%%%%%%%%
\begin{figure*}
  \centering
  \includegraphics[]{"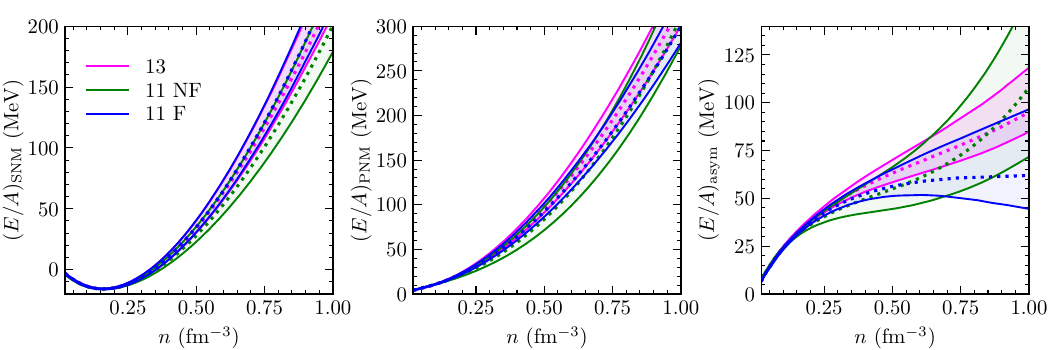"}
  \caption{Energy per nucleon in SNM (left), PNM (middle), and asymmetry energy ($\asym{E}$) (right) as functions of density.
  Medians (dotted curves) and the bounds of the 90\% CI (solid lines) for three sets of models.
  For details, see Sec.~\ref{sec:Results}.
  }
  \label{Fig:CD_E_NM}
\end{figure*}
%%%%%%%%%%%%%%%%%%%%%%%%%%%%%%%%%%%%%%%%%%%%%%%%%%%%%%%%%%%%%%%%%%%%%%

%%%%%%%%%%%%%%%%%%%%%%%%%%%%%%%%%%%%%%%%%%%%%%%%%%%%%%%%%%%%%%%%%%%%%%
\begin{figure*}
  \centering
  \includegraphics[]{"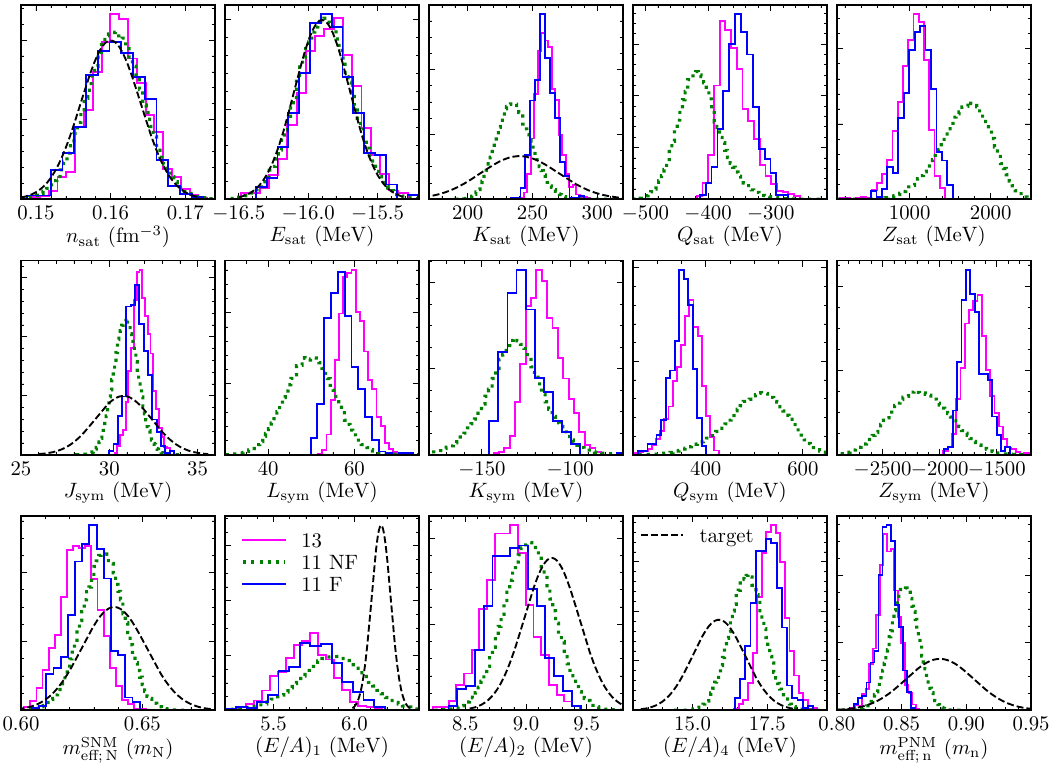"}
  \caption{Marginalized posteriors of the NEPs; the effective nucleon mass in SNM at $n=0.16~{\mathrm{fm}}^{-3}$;
  the effective neutron mass in PNM at $n=0.16~{\mathrm{fm}}^{-3}$; and energy per particle in PNM at densities $n=0.04,~0.08$, and $0.16~{\mathrm{fm}}^{-3}$. 
  When available, target distributions for constrained parameters are also plotted (black dashed curves). 
  For comparison, marginalized posteriors corresponding to run~2 in Paper~II (11 NF) and a subset of the latter consisting of models that, except for the condition on $M_{\mathrm{DU}}$, satisfy all constraints listed in Sec.~\ref{ssec:Constraints} (11 F) are also illustrated. 
  }
  \label{Fig:Hist_NM}
\end{figure*}
%%%%%%%%%%%%%%%%%%%%%%%%%%%%%%%%%%%%%%%%%%%%%%%%%%%%%%%%%%%%%%%%%%%%%%

The impact of each constraint on NM is examined by analyzing modifications in the conditional probability distributions of $\mN$, $\mn$, the energy per particle in SNM and PNM, and $\asym{E}$ as functions of density.
The results for the three sets of models discussed above are illustrated in Figs.~\ref{Fig:CD_meff_NM} and \ref{Fig:CD_E_NM}.

11\,F vs. 11\,NF: Fig.~\ref{Fig:CD_meff_NM} demonstrates that replacing $n_l$ with the significantly higher value of $n_c^*$ in (C2) excludes models with a sharp increase in $\mN(n)$.
From Eq.~\eqref{eq:meff:BSk}, it follows that $\mN(n)/m_{\mathrm{n}} \leq 1$ is equivalent to $\eff{\widetilde C} (n)\geq 0$.
Applying the latter condition up to higher densities is equivalent to favoring a stiffer behavior of $E/A$ in SNM, consistent with the results shown in the left panel of Fig.~\ref{Fig:CD_E_NM}.
The distribution of $\mN(n)$ for 11\,F is significantly narrower than for 11\,NF, and lower values of $\mN(n)$ are preferred.
The minimum of the U-shaped dependence lies in the range $3\sat{n} \lesssim n \lesssim 4\sat{n}$ and is situated between $0.5 m_{\mathrm N}$ and $0.56 m_{\mathrm N}$.
Similarly, the replacement of $n_l$ by $n_c^*$ in (C3) filters out the models with a strong decrease in $\mn(n)$, which is caused by high values of $\eff{\widetilde C}(n)+\eff{\widetilde D}(n)$, as well as those where $\mn(n)$ starts to increase at $\sat{n}-2\sat{n}$. 
The behavior of $\mn(n)$ in 11\,F ranges from smoothly decreasing with density to U-shaped.
Fig.~\ref{Fig:CD_E_NM} confirms that at high densities, the upper boundary of $E/A$ in PNM is slightly lower for 11\,F than for 11\,NF.
Also, for 11\,F, the lower boundary of $E/A$ in PNM is higher than that for 11\,NF, especially over $0.4 \lesssim n~[\mathrm{fm}^{-3}] \lesssim 0.8$.
The net effect is that the median curve for $E/A$ in PNM reaches higher values in 11\,F than in 11\,NF for densities $0.4 \lesssim n~[\mathrm{fm}^{-3}] \lesssim 0.8$. 

U-shaped behaviors for both $\mN(n)$ and $\mn(n)$ in 11\,F imply that most of the models in this set feature negative thermal pressures~\cite{Raduta_AA_2026}; 
when implemented in numerical simulations of core-collapse supernovae or binary NS mergers, these may lead to responses not present in cases where nucleonic EoSs with monotonic density dependencies of the effective masses are used.
The interplay between an increased $E/A$ in SNM and a decreased $E/A$ in PNM, especially at high densities, leads to much lower values of $\asym{E}(n)$ in 11\,F than in 11\,NF.
The shape of the curves corresponding to the median and the lower quantile indicates that for about half of the models in 11\,F, $\asym{E}(n)$ is non-monotonic. 
This suggests that $\beta$-equilibrated matter has a lower proton fraction in 11\,F than in 11\,NF.

13 vs. 11\,F: the inclusion of extra terms $t_4 x_4$ and $t_5 x_5$, intended to increase EDF flexibility, has little impact on the behavior of $\mN(n)$ and $\mn(n)$.
On the other hand, requiring the direct Urca process in NSs with masses $M \geq 1.5~\Msun$, which assumes that $Y_p \geq [1+(1+x_e^{1/3})^3]^{-1}$, results in an increase in $E/A$ for PNM and in $\asym{E}$ compared to the 11\,F set ($x_e = n_e/n_p$; $n_e$ and $n_p$ are electron and proton number densities).
Fig.~\ref{Fig:CD_E_NM} shows that the distribution of $\asym{E}(n)$ is relatively narrow for the new set of models, and for all models in the new set, $\asym{E}$ increases with $n$.
Also, $E/A$ in SNM is only slightly lower in the new set compared to that of 11\,F.

\looseness=-1
Insight into how the increased flexibility of the full 13-parameter EDF and the sets of constraints impact the NEPs and the behavior of sub-saturated PNM is provided in Fig.~\ref{Fig:Hist_NM}.
The additional constraints present in the new set and 11\,F impact all NEPs except $\sat{n}$ and $\sat{E}$.
This is no surprise given that all extra constraints are effective at high densities and the prior domains for $\sat{n}$ and $\sat{E}$ are narrow.
Marginalized posteriors of $\sat{K}$, $\sat{Q}$, and $\sat{Z}$ show that the stiffening of $E/A$ in SNM for 11\,F relative to 11\,NF is primarily due to an increase in $\sat{K}$, which is the lowest-order parameter in this list.
Similarly, marginalized posteriors of $\sym{J}$, $\sym{L}$, $\sym{K}$, $\sym{Q}$ and $\sym{Z}$ show that the (intermediate density increase) high-density decrease of $E/A$ in PNM for 11\,F compared to 11\,NF is driven by ($\sym{L}$ and $\sym{K}$) $\sym{Q}$ and $\sym{Z}$, which are the (lowest) highest order coefficients considered in the isovector expansion.
The constraints listed in Sec.~\ref{ssec:Constraints} favor stiffer behavior of PNM at sub-saturation densities.
Indeed, $(E/A)_2$ ($(E/A)_4$) is lower (higher) in 11\,F relative to 11\,NF, which means extra tension with the target distributions.
Also, values of $\mn$ and $\mN$ at $0.16~\mathrm{fm}^{-3}$ are lower in 11\,F compared to 11\,NF, which indicates poorer agreement with the target distributions.
With the exception of $\sym{K}$ and, to a lesser extent, $\sym{L}$, marginalized posteriors of the new set are almost identical to those of the 11\,F set.
These findings confirm the discussion in Appendix~A of Paper~II, where $x_5=0$ was conjectured to affect the density dependence of the symmetry energy, while the isoscalar sector should be insensitive to both $x_4=0$ and $x_5=0$.

For the medians, upper, and lower boundaries of the 90\% confidence intervals (CIs) of the NEPs, $\mN$, and $\mn$ at $0.16~\mathrm{fm^{-3}}$, see Table~\ref{tab:Prop} in Appendix~\ref{App:TableProp}.

%%%%%%%%%%%%%%%%%%%%%%%%%%%%%%%%%%%%%%%%%%%%%%%%%%%%%%%%%%%%%%%%%%%%%%
\subsection{Neutron star matter}
\label{ssec:NS}
%%%%%%%%%%%%%%%%%%%%%%%%%%%%%%%%%%%%%%%%%%%%%%%%%%%%%%%%%%%%%%%%%%%%%%

%%%%%%%%%%%%%%%%%%%%%%%%%%%%%%%%%%%%%%%%%%%%%%%%%%%%%%%%%%%%%%%%%%%%%%
\begin{figure}
  \centering
  \includegraphics[]{"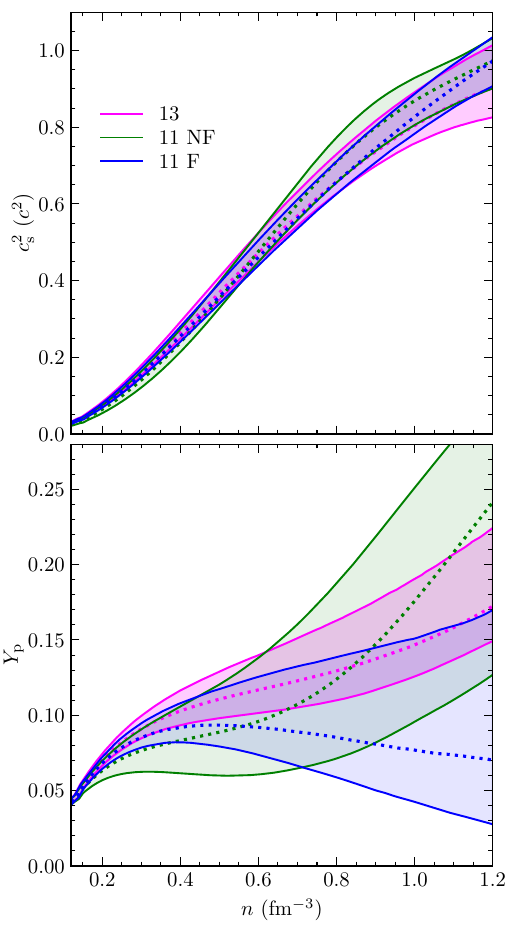"}
  \caption{
    Speed of sound squared (top) and proton fraction (bottom) in NS matter as functions of density.
    Medians (dotted curves) and the bounds of the 90\% CI (solid lines) for three sets of models.
    For details, see Sec.~\ref{sec:Results}.
  }
  \label{Fig:CD_NS}
\end{figure}
%%%%%%%%%%%%%%%%%%%%%%%%%%%%%%%%%%%%%%%%%%%%%%%%%%%%%%%%%%%%%%%%%%%%%%%

%%%%%%%%%%%%%%%%%%%%%%%%%%%%%%%%%%%%%%%%%%%%%%%%%%%%%%%%%%%%%%%%%%%%%%
\begin{figure*}
  \centering
  \includegraphics[]{"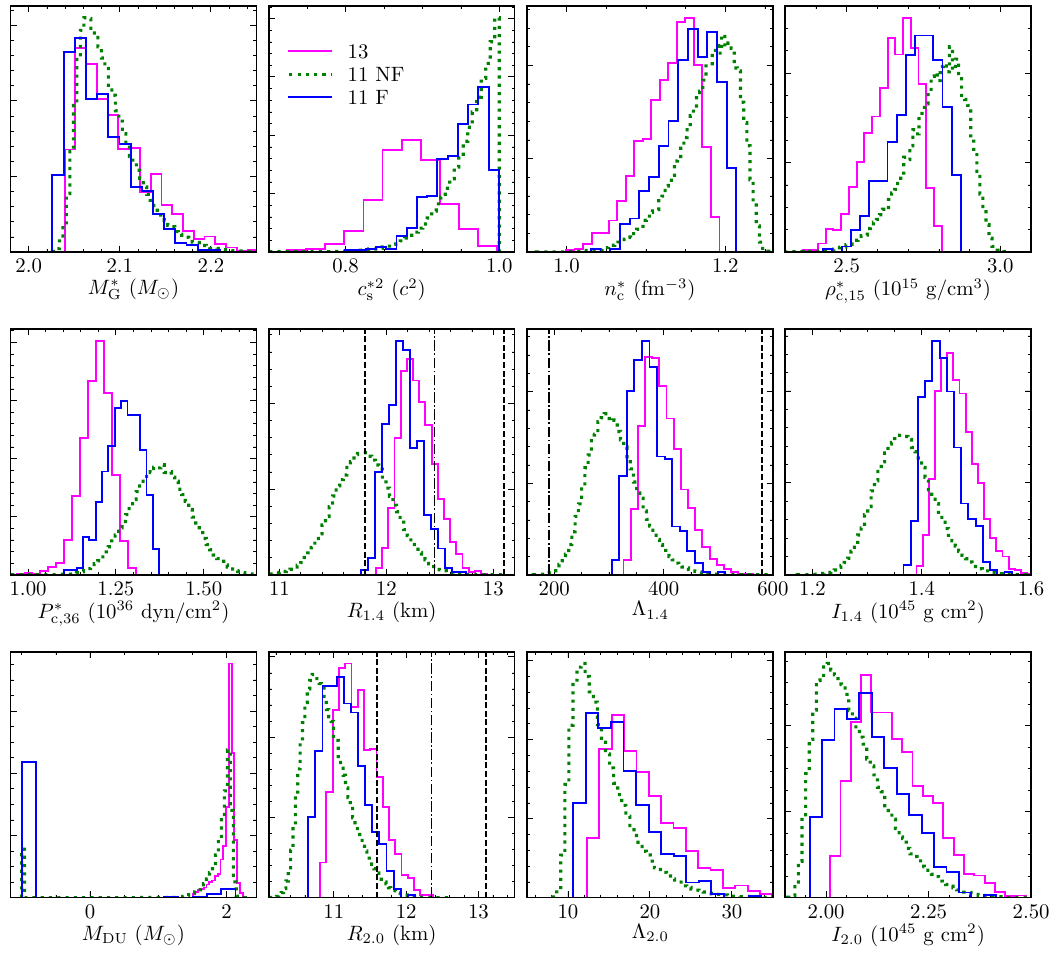"}
  \caption{Marginalized posteriors of selected properties of NSs. 
    The plotted quantities include: the maximum gravitational mass ($M_{\rm G}^*$); the central density corresponding to the most massive configuration ($n_{\mathrm c}^*$); the speed of sound squared ($c_{\rm s}^{*2}$), energy density ($\rho_{\rm c}^*$), and pressure ($P_{\rm c}^*$) at $n_{\mathrm c}^*$; radii ($R_{1.4}$, $R_{2.0}$), tidal deformabilities ($\Lambda_{1.4}$, $\Lambda_{2.0}$), and moments of inertia  ($I_{1.4}$, $I_{2.0}$) of NSs with masses equal to $1.4~\Msun$ and $2.0~\Msun$; and the minimum NS mass that allows for the direct Urca process ($M_{\rm DU}$). 
    The vertical dot-dashed and dashed lines on the $R_{1.4}$ and $R_{2.0}$ plots illustrate the median and the lower/upper bounds (at 68\% confidence), respectively, of the constraints $R=12.45 \pm 0.65~\mathrm{km}$ and $R=12.35 \pm 0.75~\mathrm{km}$ from Ref.~\cite{Miller_ApJL_2021}.
    The vertical dot-dashed and dashed lines on the $\Lambda_{1.4}$ plot illustrate the median and the upper bound (at 90\% confidence), respectively, of the constraint $\Lambda_{1.4} = 190^{+390}_{-120}$  from Ref.~\cite{Abbott_PRL_121}.
    For $M_{\rm DU}$, the value ``$-1$'' corresponds to models that do not allow this process to operate in stable NSs.
    The same set of models as in Fig.~\ref{Fig:Hist_NM} is considered.
 }
  \label{Fig:Hist_NS}
\end{figure*}
%%%%%%%%%%%%%%%%%%%%%%%%%%%%%%%%%%%%%%%%%%%%%%%%%%%%%%%%%%%%%%%%%%%%%%

We build NS EoSs by smoothly matching the core and crust EoSs at a density of about $\sat{n}/2$.
For the outer and inner crusts, we adopt the models by Haensel, Zdunik and Dobaczewski~\cite{HDZ_1989} and Negele and Vautherin~\cite{NV_1973}, respectively.
The core EoS is computed by solving the $\beta$-equilibrium, $\mu_n=\mu_p+\mu_e$, and net charge neutrality equations, $n_p = n_e + n_{\mu}$.
For neutrino-transparent matter, chemical equilibrium with respect to weak interactions implies $\mu_e=\mu_{\mu}$.

Figure~\ref{Fig:CD_NS} depicts the density dependence of the speed of sound squared ($c_S^2/c^2=dP/de$) and proton fraction ($Y_p=n_p/n$) of $\beta$-equilibrated matter.
A comparison of the upper bounds of the 90\% CI for 11\,NF and 11\,F in the top panel reveals that requiring causality for both NS matter and PNM up to $n_c^*$, which exceeds the value of $n_l$ by almost 50\%, eliminates models with high $c_S$ values at $n \gtrsim 4\sat{n}$.
A comparison of the lower bounds of the 90\% CI for 11\,NF and 11\,F shows that models with low values of $c_S$ for $n \lesssim 3\sat{n}$ are also excluded.
All in all, with the exception of the highest densities considered in Fig.~\ref{Fig:CD_NS}, which coincide with the peak of the $n_c^*$ distribution in 11\,NF and exceed the corresponding peak in 11\,F, models in 11\,F feature lower values of $c_S^2$ compared to models in 11\,NF and, of course, less dispersion.
The extra flexibility of the 13-parameter EDF widens the range of behaviors in the sense that more models with low values of $c_S^2$ for $n \gtrsim 1~\mathrm{fm^{-3}}$ are allowed.

\looseness=-1
As expected based on the behavior of $\asym{E}(n)$ illustrated in Fig.~\ref{Fig:CD_E_NM} (right panel), the three classes of models feature very different proton fractions in NS matter.
Models in 11\,NF, where $\asym{E}$ increases with $n$, show that in the majority of cases, $Y_p$ also increases with $n$.
Models in 11\,F, which feature a moderate increase or even non-monotonic behavior of $\asym{E}(n)$, manifest a completely different pattern.
First, the number of models where $Y_p(n)$ has a rise-and-fall behavior exceeds the number of models where $Y_p$ monotonically increases, as evidenced by the bell-shaped median curve.
The fact that the lower quantile of models in 11\,NF bears no signature of the rise-and-fall behavior means that these models represent less than 5\% of the models in 11\,NF.
Then, for $n \gtrsim 0.7~\mathrm{fm}^{-3}$, models in 11\,F lead to much lower values of $Y_p$ than models in 11\,NF.
This suggests that the number of models that accommodate the direct Urca process will be smaller in 11\,F than in 11\,NF.
Finally, for $n \lesssim 0.45~\mathrm{fm}^{-3}$, models in 11\,F have higher values of $Y_p$ than models in 11\,NF.
The latter two features could have been anticipated based on the dependence of $\asym{E}(n)$ illustrated in Fig.~\ref{Fig:CD_E_NM} (right panel).
Models in the new set feature a $Y_p$ that increases with $n$ and show much less dispersion in $Y_p(n)$.
These are the consequences of the condition $M_{\mathrm{DU}} \ge 1.5~\Msun$, which filters out models with low values of $\asym{E}$ at high densities.
The maximum values allowed by this set of models at densities in excess of $0.7~\mathrm{fm}^{-3}$ are much lower than the corresponding values for the 11\,NF set and considerably higher than those for the 11\,F set.

Further insight into the properties of NSs is provided in Fig.~\ref{Fig:Hist_NS} in terms of the marginalized posteriors of some selected global properties.
With the exception of $M_{\mathrm G}^*$, all quantities are affected by the flexibility of the EDF and the specific constraints, indicating a high sensitivity to $\sym{L}$ and $\sym{K}$.
The exclusion of models with high $c_S^2$ values at high densities in the 11\,F set (relative to 11\,NF) is reflected in the $c_S^{*2}/c^2$ histogram, which is no longer peaked at unity, and in the $n_{\mathrm{c}}^*$, $\rho_{\mathrm{c}}^*$, and $P_{\mathrm{c}}^*$ histograms, which are shifted to the left. 
The latter are typical diagnostics for stiff EoSs, as confirmed by the larger values of $R_{1.4}$, $\Lambda_{1.4}$, $I_{1.4}$, $R_{2.0}$, $\Lambda_{2.0}$, and $I_{2.0}$. 
The new set of models features even stiffer EoSs.
We notice that for the models built here as well as those in 11\,F, the lowest values of $R_{1.4}$, $\Lambda_{1.4}$, and $I_{1.4}$ are significantly higher than those in 11\,NF.
This is because models with low values of $c_S^2$ at low densities are filtered out by the updated set of constraints.

\looseness=-1
The values of central density, energy density, and pressure of the maximum mass configuration are lower than those in the 11\,F set, while the values of $R_{1.4}$, $\Lambda_{1.4}$, $I_{1.4}$, $R_{2.0}$, $\Lambda_{2.0}$, and $I_{2.0}$ are higher than those obtained in 11\,F.
Marginalized posteriors of radii ($R_{1.4}$ and $R_{2.0}$) and tidal deformability ($\Lambda_{1.4}$) are compared with the constraints from Refs.~\cite{Miller_ApJL_2021} and \cite{Abbott_PRL_121}, respectively (shown by dashed and dot-dashed vertical lines).
By favoring stiffer models, the updated set of constraints reduces the tension with the data in Ref.~\cite{Miller_ApJL_2021} for both $R_{1.4}$ and $R_{2.0}$. 
The $R_{1.4}$ histograms of these sets lie between the lower and upper quantiles (at 68\% CI) from Ref.~\cite{Miller_ApJL_2021}, while for $R_{2.0}$, most of our models provide values lower than the lower quantile (at 68\% CI) from Ref.~\cite{Miller_ApJL_2021}.
These results suggest that even the models in 11\,F and the new set are too soft.
However, a comparison of $\Lambda_{1.4}$ with the results of Ref.~\cite{Abbott_PRL_121} conveys the opposite message: our models are too stiff; the 11\,F set and the new set are in stronger tension with the data than the models in 11\,NF.
This discrepancy may reflect a tension between the conclusions of Refs.~\cite{Miller_ApJL_2021} and \cite{Abbott_PRL_121}, or an inability of the Brussels-Skyrme EDF to describe the data.
In either case, this finding might contribute to a better understanding of the EoS issue and deserves further investigation.

Also, for the new set of models, the $c_S^{*2}/c^2$ distribution peaks at a density of $0.85-0.9$~fm$^{-3}$ and is almost symmetric.
By construction, all these models allow for the direct Urca process. 
However, Fig.~\ref{Fig:Hist_NS} shows that in most cases the direct Urca threshold mass is around $2~\Msun$.
The 11\,NF set allows both for models where the direct Urca process is forbidden in stable stars and for models that support it. 
As with the new set, the marginalized posterior is peaked around $2~\Msun$.
For 11\,F, only a small fraction of models allow for the direct Urca process, and the DU threshold mass is $\approx 2~\Msun$. 

The moment of inertia is calculated up to the leading order in the slow, rigid rotation approximation~\cite{Hartle_ApJ_1967}.
The corresponding panels in Fig.~\ref{Fig:Hist_NS} show that the marginalized posteriors of $I_{1.4}$ and $I_{2.0}$ replicate the marginalized posteriors of $R_{1.4}$ and $R_{2.0}$.

For the medians and the upper and lower bounds of the 90\% CI for the NS global parameters, see Table~\ref{tab:Prop} in Appendix~\ref{App:TableProp}.

%%%%%%%%%%%%%%%%%%%%%%%%%%%%%%%%%%%%%%%%%%%%%%%%%%%%%%%%%%%%%%%%%%%%%%
\subsection{Correlations}
\label{ssec:Correl}
%%%%%%%%%%%%%%%%%%%%%%%%%%%%%%%%%%%%%%%%%%%%%%%%%%%%%%%%%%%%%%%%%%%%%%

%%%%%%%%%%%%%%%%%%%%%%%%%%%%%%%%%%%%%%%%%%%%%%%%%%%%%%%%%%%%%%%%%%%%%%
\begin{figure*}
  \centering
  \includegraphics[]{"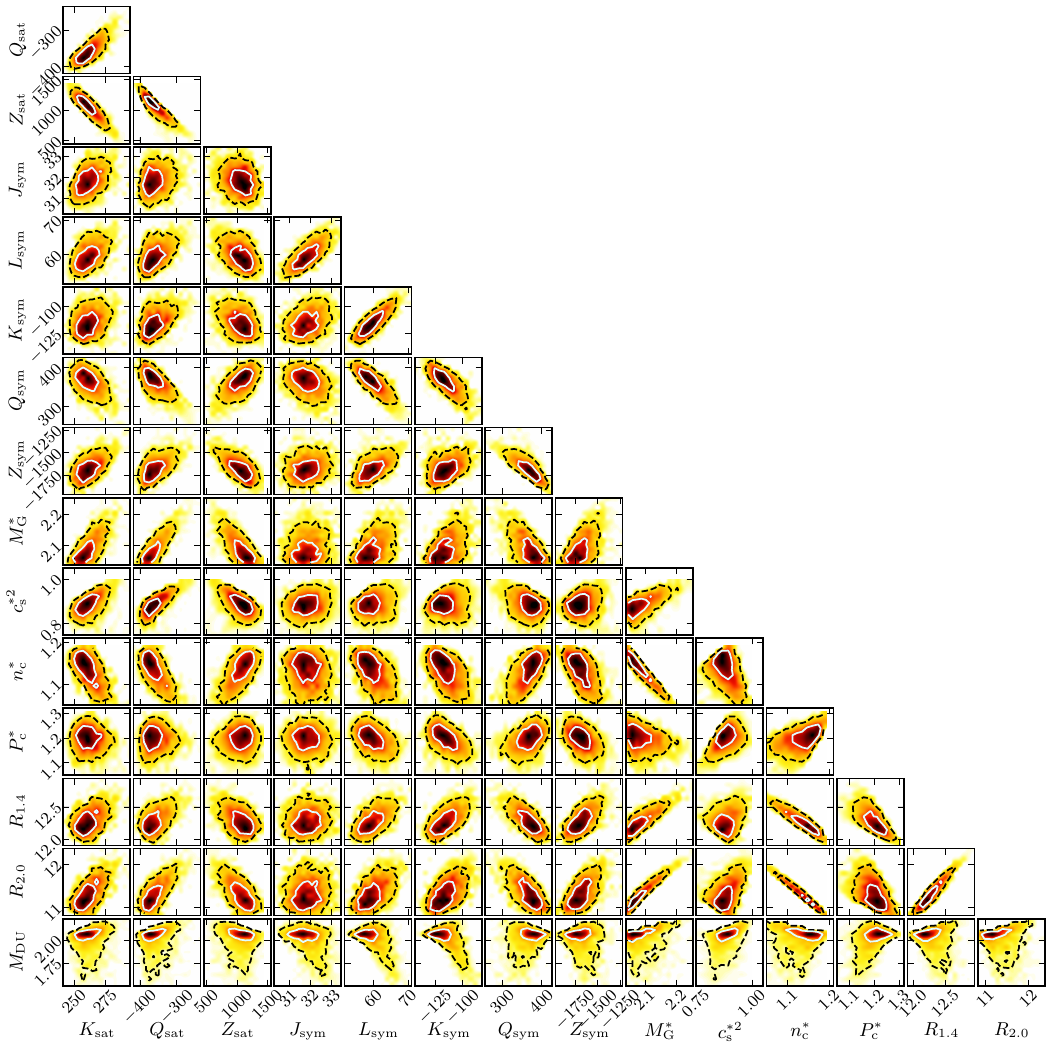"}
  \caption{
    Two-dimensional (2D) marginalized posteriors of selected NM and NS parameters.
    The solid and dashed contours show 50\% and 90\% CR, respectively. 
    The results correspond to the models built in this work.
  }
  \label{Fig:Corner_NS}
\end{figure*}
%%%%%%%%%%%%%%%%%%%%%%%%%%%%%%%%%%%%%%%%%%%%%%%%%%%%%%%%%%%%%%%%%%%%%%

Numerous phenomenological and agnostic models were employed to systematically investigate correlations between global properties of NSs and NEPs.
This effort was necessary to understand the sensitivity of NS properties to properties of NM and to distinguish physical correlations, which manifest in all models where the quantities under consideration span sufficiently wide domains, from spurious ones.
The latter stem from particularities of the model and are more frequent in phenomenological models than in agnostic ones.

In Fig.~\ref{Fig:Corner_NS}, we investigate a series of such correlations for the EoS models built in this work.
No straightforward comparison with the results of models in 11\,NF is possible, as Paper~II shows correlation plots for runs other than run~2.
According to Fig.~\ref{Fig:Corner_NS}: 
i) correlations are observed between NEPs in the isoscalar sector: $\sat{Z}$ is negatively correlated with $\sat{K}$ and $\sat{Q}$, while a positive correlation exists between $\sat{K}$ and $\sat{Q}$.
None of these is particularly strong, which might be due to the increased flexibility of the EDF, narrow domains for each quantity, or both;
ii) correlations are also observed between NEPs in the isovector sector: the correlations $\sym{J}-\sym{L}$ and $\sym{L}-\sym{K}$ are positive, while the correlations $\sym{L}-\sym{Q}$, $\sym{K}-\sym{Q}$, and $\sym{Q}-\sym{Z}$ are negative, and all of them are weaker than those in the isoscalar channel;
iii) no significant correlations are observed between $\sat{X}$ and $\sym{Y}$, where $X=K,Q,Z$ and $Y=J,L,K$; 
iv) $M_{\mathrm{G}}^*$ is positively (negatively) correlated with $\sat{K}$ and $\sat{Q}$ ($\sat{Z}$);
v) $M_{\mathrm{G}}^*$ is positively (negatively) correlated with $c_S^{2*}$, $R_{1.4}$, and $R_{2.0}$ ($n_c^*$);
vi) $n_c^*$ is negatively correlated with $R_{1.4}$ and $R_{2.0}$, while the correlation $R_{1.4}-R_{2.0}$ is positive and strong;
vii) $R_{1.4}$ and $R_{2.0}$ are weakly correlated with $\sym{L}$, $\sym{K}$, and $\sym{Z}$ (positively) and with $\sym{Q}$ (negatively).
Figure~6 in Paper~II, which corresponds to run~3 in that paper and is more constrained than the 11\,NF set, features the same correlations as Fig.~\ref{Fig:Corner_NS}, although with different strengths. 
This means that the extra flexibility of the 13-parameter EDF does not wash out correlations present in the 11-parameter model.

We notice that $M_{\mathrm{DU}}$ does not appear to be correlated with any NEPs.
This result is in contrast with the findings in Fig.~11 of Paper~I, where standard Skyrme interactions were employed, and in accordance with the results presented in Figs.~7 and 8 of Paper~II, where Brussels extended Skyrme interactions with $x_4=x_5=0$  were used.
The fact that the negative correlations $M_{\mathrm{DU}} - \sym{L}$ and $M_{\mathrm{DU}} - \sym{K}$ found in Paper~I are washed out when the EDF is given more flexibility suggests that these correlations are not physical.

%%%%%%%%%%%%%%%%%%%%%%%%%%%%%%%%%%%%%%%%%%%%%%%%%%%%%%%%%%%%%%%%%%%%%%
\subsection{Thermal response}
\label{ssec:FiniteT}
%%%%%%%%%%%%%%%%%%%%%%%%%%%%%%%%%%%%%%%%%%%%%%%%%%%%%%%%%%%%%%%%%%%%%%

%%%%%%%%%%%%%%%%%%%%%%%%%%%%%%%%%%%%%%%%%%%%%%%%%%%%%%%%%%%%%%%%%%%%%%
\begin{figure*}
  \centering
  \includegraphics[]{"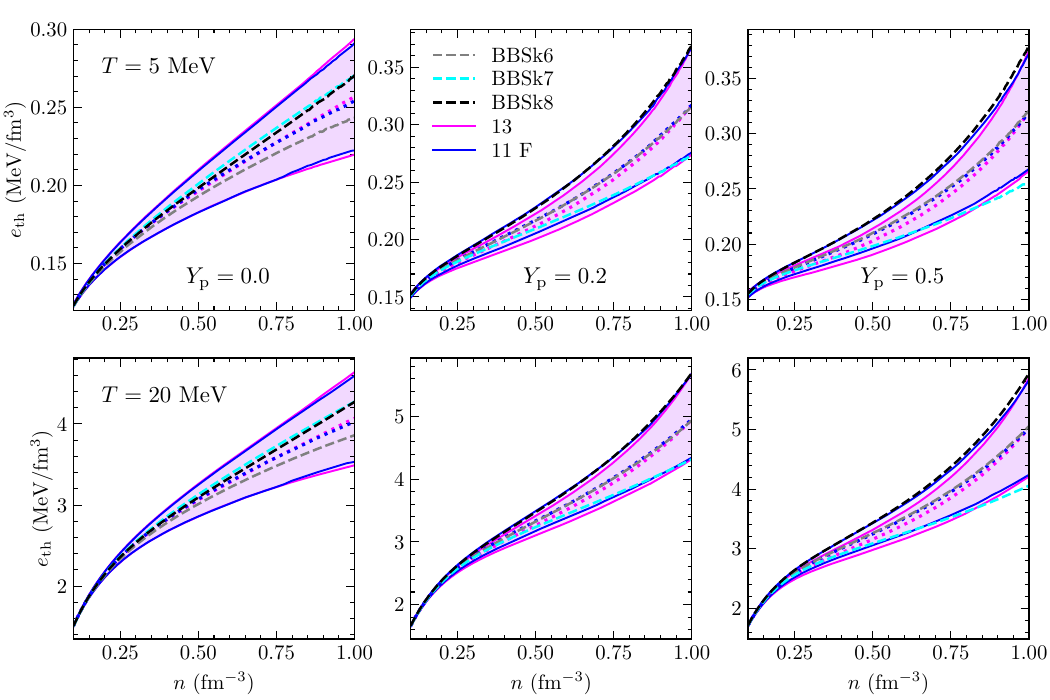"}
  \caption{Thermal energy (Eq.~\eqref{eq:eth}) as a function of density for NM with $Y_p=0,\,0.2$, and $0.5$ at $T=5$~MeV (top panels) and 20~MeV (bottom panels).
   In addition to the medians and the upper and lower bounds of the 90\% CI corresponding to the current set of models and 11~F, we also show predictions from BBSk6, BBSk7, and BBSk8 EoSs from Paper~II.
  }
  \label{Fig:eth}
\end{figure*}
%%%%%%%%%%%%%%%%%%%%%%%%%%%%%%%%%%%%%%%%%%%%%%%%%%%%%%%%%%%%%%%%%%%%%%

%%%%%%%%%%%%%%%%%%%%%%%%%%%%%%%%%%%%%%%%%%%%%%%%%%%%%%%%%%%%%%%%%%%%%%
\begin{figure*}
  \centering
  \includegraphics[]{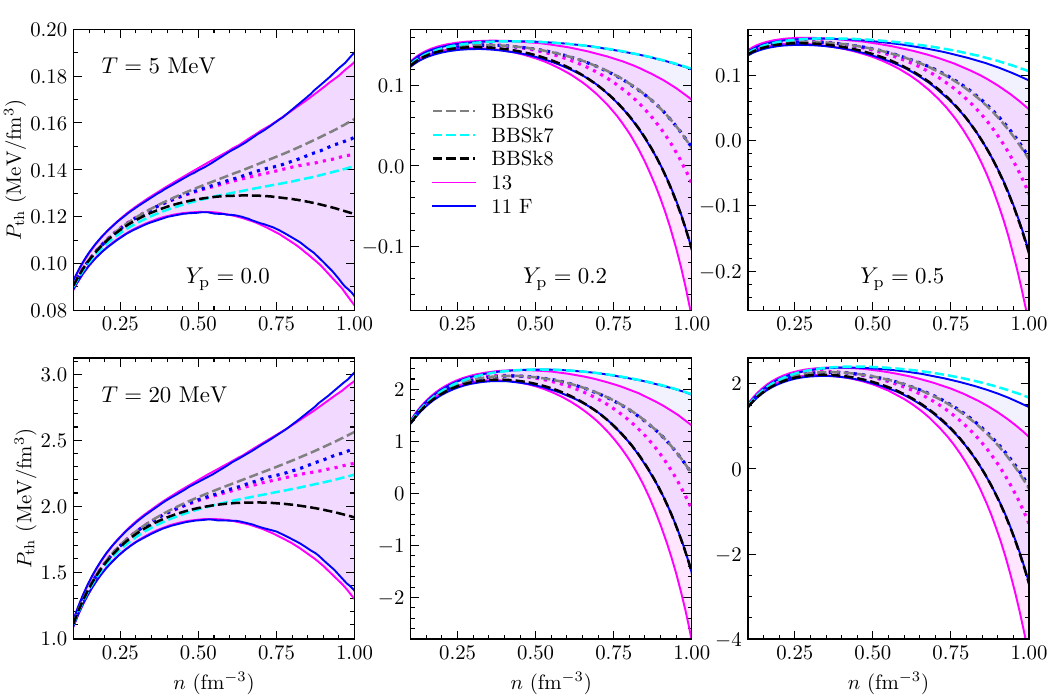}
  \caption{Same as Fig.~\ref{Fig:eth}, but showing the thermal pressure given by Eq.~\eqref{eq:pth}.
  }
  \label{Fig:pth}
\end{figure*}
%%%%%%%%%%%%%%%%%%%%%%%%%%%%%%%%%%%%%%%%%%%%%%%%%%%%%%%%%%%%%%%%%%%%%%

Following Refs.~\cite{Constantinou_PRC_2014,Constantinou_PRC_2015,Raduta_PLB_2024,Raduta_AA_2026}, we investigate the finite-temperature behavior of NM in terms of $\tth{e}$ and $\tth{p}$ as functions of density for fixed values of $Y_p$ and temperature.
In addition to the EoSs generated in this work and EoSs in 11\,F set, we also show the behavior of BBSk6, BBSk7, and BBSk8 EoSs.
The latter effective interactions were generated in Paper~II. 
General-purpose EoS tables built for these interactions according to Ref.~\cite{Raduta_AA_2025} are publicly available in the \textsc{CompOSE}~\cite{Typel_EPJA_2022} online repository\,\footnote{\url{https://compose.obspm.fr/}}.
For their characterization in terms of NEPs and the values of global NS properties, see Table~\ref{tab:Prop}. 

Figure~\ref{Fig:eth} shows the behavior of $\tth{e}$ for $Y_p=0$, 0.2, and 0.5 at $T= 5$ and 20~MeV.
In all considered cases, $\tth{e}$ increases with density and temperature.
In most cases, for a given temperature and density, higher values of $Y_p$ lead to higher values of $\tth{e}$.
However, this is not a universal rule; see, e.g., BBSk7.
BBSk7 has lower values of $m_{\mathrm{eff};n/p}$ compared to those of BBSk6 and BBSk8 for $Y_p=0.2$ and 0.5; also, for BBSk7, $m_{\mathrm{eff};n/p}(n)$ has the  least pronounced U-shape (not shown).
Finally, the EoSs generated in this work behave similarly to those in the 11\,F set. 
This was expected based on Eq.~\eqref{eq:eth} and the similarity between the effective masses for these two sets of models; see Fig.~\ref{Fig:CD_meff_NM}.

Figure~\ref{Fig:pth} demonstrates the behavior of $\tth{p}(n)$ for the temperatures and proton fractions considered in Fig.~\ref{Fig:eth}.
For $n \lesssim 3 \sat{n}$, the dispersion among the models is small and, in all cases,  $\tth{p}$ increases with density.
For $n \gtrsim 3 \sat{n}$, both the EoSs generated here and those in 11\,F exhibit a wide range of behaviors.
For $Y_p=0$, more (less) than half of the models built here (in 11\,F) predict that $\tth{p}$ increases with density. 
For $Y_p=0.2$ and 0.5, all EoSs exhibit a non-monotonic ``rise-and-fall'' behavior of $\tth{p}(n)$. 
The same holds true for $Y_p=0.02$ (not shown).
The least (most) pronounced rise-and-fall behavior of $\tth{p}(n)$ corresponds to BBSk6 (BBSk8), which has the least (most) pronounced U-shaped density dependence of $m_{\mathrm{eff};i}$.
A significant number of models predict that, at high densities, $\tth{p}<0$.
As explained at length in Ref.~\cite{Raduta_AA_2026}, this is a consequence of the U-shaped $m_{\mathrm{eff};i}(n)$.
The differences between the quantiles (median, upper, and lower) of the models developed in this work and those in 11\,F are more pronounced than the difference shown in Fig.~\ref{Fig:eth}, because, unlike $\tth{e}$, $\tth{p}$ also depends on $\partial m_{\mathrm{eff};i}/\partial n$ (see Eq.~\eqref{eq:pth}).

%%%%%%%%%%%%%%%%%%%%%%%%%%%%%%%%%%%%%%%%%%%%%%%%%%%%%%%%%%%%%%%%%%%%%%
\section{Conclusion}
\label{sec:Concl}
%%%%%%%%%%%%%%%%%%%%%%%%%%%%%%%%%%%%%%%%%%%%%%%%%%%%%%%%%%%%%%%%%%%%%%

In this work, we have extended and generalized the studies presented in Paper~II concerning the ability of non-relativistic models with Brussels extended Skyrme interactions to describe matter at densities exceeding several times $\sat{n}$ and isospin asymmetries $0 \leq \delta \leq 1$.
To fully utilize the flexibility of the EDF, all 13 parameters of the effective interaction were allowed to vary.
To ensure physical consistency across the density range relevant for NSs (assuming only nucleonic degrees of freedom), ``sanity'' conditions on the NS and PNM EoSs were imposed up to the central density of the maximum mass configuration ($n_c^*\approx 1.0-1.2~\mathrm{fm}^{-3}$).
These conditions included thermodynamic stability, causality, and constraints on neutron and proton effective masses and Fermi velocities.
The parameter space was explored via an MCMC procedure, with models weighted based on their agreement with fiducial values for the four best-known NEPs ($\sat{n}$, $\sat{E}$, $\sat{K}$, and $\sym{J}$), $\chi$EFT data for the energy per particle in PNM and neutron effective mass in PNM and SNM (all within $0.08 \leq n \, [\mathrm{fm}^{-3}] \leq 0.16$), as well as the requirements that $M_\mathrm{G}^* > 2~\Msun$ and that the electron direct Urca threshold mass be $M_\mathrm{DU} \ge 1.5~\Msun$.

The tighter constraints employed in this work, compared to those in Paper~II, filter out a large number of models with extreme behaviors of $\mn(n)$, $\mN(n)$, and $\asym{E}(n)$, leading to significant modifications of the marginalized posterior distributions of all NS global parameters.
NS EoSs become stiffer, leading to larger values of $R_{1.4}$, $\Lambda_{1.4}$, $I_{1.4}$, $R_{2.0}$, $\Lambda_{2.0}$, and $I_{2.0}$ compared to those in Paper~II.
While the extra flexibility of the 13-parameter EDF further stiffens the EoS, agreement with the constraints from Refs.~\cite{Miller_ApJL_2021} and \cite{Abbott_PRL_121} remains modest, likely because the inherent flexibility of the phenomenological EDF is still limited.
It is interesting to notice that, with the exception of $\sym{L}$ and $\sym{K}$, the NEPs are unable to capture the way in which the behavior of NM is modified when increasing the number of effective interaction parameters from 11 to 13.

For all the models built here, $\mN(n)$ is U-shaped, and in most cases, $\mn(n)$ is also U-shaped. 
These features result in a pronounced ``rise-and-fall'' behavior of $\tth{p}(n)$ in NM with $Y_p \gtrsim 0.01$ and negative thermal pressures at high densities. 
We expect that these EoSs, when implemented in numerical simulations of core-collapse supernovae and binary NS mergers, will produce signatures distinct from those obtained using nucleonic models with standard Skyrme interactions.

Alongside other studies in the literature, e.g., Refs.~\cite{Papakonstantinou_PRC_2023,Imam_PRD_2024,Beznogov_ApJ_2024,Beznogov_PRC_2024}, which employ nonrelativistic mean-field models, and Refs.~\cite{Traversi_ApJ_2020,Malik_ApJ_2022,Beznogov_PRC_2023,Malik_PRD_2023,Char_PRD_2023,Imam_PRD_2024,Li_PRC_2025,Cartaxo_ApJSS_2026}, which employ relativistic mean-field models, this work aims to determine the extent to which phenomenological models are adequate for describing dense matter, and how present and future constraints from multi-messenger astronomy of NSs can improve our understanding of NM.  
 
\vspace{0.5cm}
Data availability: in Ref.~\cite{BSk_13Par_data} we provide the posterior distributions of the input parameters and the corresponding NEPs for the EoS models in our main (``13'') run.

%%%%%%%%%%%%%%%%%%%%%%%%%%%%%%%%%%%%%%%%%%%%%%%%%%%%%%%%%%%%%%%%%%%
\begin{acknowledgments}
  We acknowledge support from the Ministry of Research, Innovation and Digitization via Project No. PN-IV-P1-PCE-2023-0324 (CNCS/CCCDI – UEFISCDI)
  and PN 23 21 01 02.

  M.V.B. and A.R.R. contributed equally to this work.
\end{acknowledgments}
%%%%%%%%%%%%%%%%%%%%%%%%%%%%%%%%%%%%%%%%%%%%%%%%%%%%%%%%%%%%%%%%%%%

%%%%%%%%%%%%%%%%%%%%%%%%%%%%%%%%%%%%%%%%%%%%%%%%%%%%%%%%%%%%%%%%%%%
\appendix
\section{Properties of NM and NSs}
\label{App:TableProp}
%%%%%%%%%%%%%%%%%%%%%%%%%%%%%%%%%%%%%%%%%%%%%%%%%%%%%%%%%%%%%%%%%%%

Table~\ref{tab:Prop} summarizes basic properties of NM and NS in terms of medians and the lower and upper bounds of the 90\% CI.
This information complements the marginalized distributions of the same quantities, illustrated in 
Figs.~\ref{Fig:Hist_NM} and \ref{Fig:Hist_NS}.

%%%%%%%%%%%%%%%%%%%%%%%%%%%%%%%%%%%%%%%%%%%%%%%%%%%%%%%%%%%%%%%%%%%%%%
\renewcommand{\arraystretch}{1.5}
\setlength{\tabcolsep}{3.9pt}
\begin{table*}
\label{tab:Prop}
\centering
\caption{Key properties of NM and NSs. We provide medians and bounds of 90\% CIs.
	The data in columns 3-4 (5-6) refer to the ensemble of models built in this work (run~2 in Paper~II).
	The data in columns 7-8 refer to models in run~2 in Paper~II that pass all the conditions in Sec.~\ref{ssec:Constraints} except the one on $M_{\mathrm{DU}}$.
	The data in columns 9, 10, and 11 refer to BBSk6, BBSk7, and BBSk8, respectively. 
	For these interactions, general-purpose EoS tables are made available on \textsc{CompOSE}.
	For NM, the table lists the saturation density ($\sat{n}$) of the SNM; the energy per nucleon ($\sat{E}$), compression modulus ($\sat{K}$),
	skewness ($\sat{Q}$), and kurtosis ($\sat{Z}$) of the SNM at $\sat{n}$; 
	the symmetry energy ($\sym{J}$), its slope ($\sym{L}$),
	compressibility ($\sym{K}$), skewness ($\sym{Q}$), and kurtosis ($\sym{Z}$) at $\sat{n}$; the Landau effective mass of the nucleons in the SNM ($m_{\mathrm{eff;\,N}}^\mathrm{SNM}$) and that of the neutrons in the PNM ($m_{\mathrm{eff;\,n}}^\mathrm{PNM}$) at 0.16~fm$^{-3}$.
	For NSs, the table lists the maximum gravitational ($M_{\mathrm{G}}^*$) and baryonic ($M_{\mathrm{B}}^*$) masses;
	the central particle density corresponding to the maximum mass configuration ($n_{\mathrm{c}}^*$);
	the speed of sound squared ($c^{*2}_{\mathrm{s}}$), energy density ($\rho_{\mathrm{c}}^*$), and pressure ($P_{\mathrm{c}}^*$) at $n_{\mathrm{c}}^*$; and the
	radii ($R_{1.4}$, $R_{2.0}$) and tidal deformabilities ($\Lambda_{1.4}$, $\Lambda_{2.0}$) of NSs with masses equal to $1.4~\Msun$ and $2.0~\Msun$.
}
\begin{tabular}{lcccccccccc}
\toprule
\toprule
	\multirow{2}{*}{Par.}               &  \multirow{2}{*}{Units}       & \multicolumn{2}{c}{13}                  & \multicolumn{2}{c}{11\,NF}    & \multicolumn{2}{c}{11\,F} & BBSk6 & BBSk7 & BBSk8  \\
\cmidrule(lr){3-4}
\cmidrule(lr){5-6}
\cmidrule(lr){7-8}
                &                         & Med.     & 90\% CI              & Med.     & 90\% CI & Med.     & 90\% CI \\
\midrule
$n_\mathrm{sat}$                           &   $\mathrm{fm^{-3}}$      & $0.161$ & $^{+0.006}_{-0.0058}$ & $ 0.161$ & $^{+0.0063}_{-0.0063}$ & $0.161$  & $^{+ 0.0055}_{-0.0065}$ & $0.167$ & $0.157$ & $0.161$ \\
$E_\mathrm{sat}$                           &   $\mathrm{MeV}$	     & $-15.8$ & $^{+0.32}_{-0.33}$   & $-15.9$  & $^{+0.33}_{-0.33}$     & $ -15.9$ & $^{+0.36}_{-0.3}$      & $-15.9$ & $-15.9$ & $-15.7$ \\
$K_\mathrm{sat}$                           &   $\mathrm{MeV}$	     & $ 261$  & $^{+15}_{-11}$      & $236$    & $^{+24}_{-20}$         & $259$   & $^{+14}_{-12}$         & $258$   & $251$   & $258$ \\
$Q_\mathrm{sat}$                           &   $\mathrm{MeV}$	     & $-361$  & $^{+55}_{-31}$      & $-415$   & $^{+64}_{-51}$         & $-349$  & $^{+43}_{-36}$         & $-364$  & $-369$  & $-346$ \\
$Z_\mathrm{sat}$                           &   $\mathrm{MeV}$	     & $1050$  & $^{+220}_{-330}$    & $1710$   & $^{+460}_{-580}$        & $1120$  & $^{+260}_{-320}$       & $1176$  & $1287$  & $1095$ \\
$J_\mathrm{sym}$                           &   $\mathrm{MeV}$	     & $31.8$  & $^{+0.93}_{-0.81}$  & $30.9$   & $^{+1.2}_{-1.1}$        & $31.6$  & $^{+0.95}_{-0.86}$      & $32.3$ & $31.8$   & $31.5$ \\
$L_\mathrm{sym}$                           &   $\mathrm{MeV}$	     & $59.6$  & $^{+5.8}_{-4.2}$    & $49.8$   & $^{+9.4}_{-9.3}$        & $56.8$  & $^{+6.3}_{-4.3}$       & $58.9$ & $58.4$   & $58.4$ \\
$K_\mathrm{sym}$                           &   $\mathrm{MeV}$	     & $-115$  & $^{+18}_{-13}$      & $-131$   & $^{+27}_{-27}$         & $-127$  & $^{+21}_{-12}$        & $-125$  & $-127$   & $-119$ \\
$Q_\mathrm{sym}$                           &   $\mathrm{MeV}$	     & $365$   & $^{+36}_{-53}$      & $507$    & $^{+90}_{-120}$        & $353$   & $^{+28}_{-43}$        & $368$   & $325$    & $337$ \\
$Z_\mathrm{sym}$                           &   $\mathrm{MeV}$	     & $-1680$ & $^{+200}_{-160}$    & $-2170$  & $^{+430}_{-410}$       & $-1720$  & $^{+180}_{-130}$      & $-1757$ & $-1603$ & $-1686$ \\
$m_{\mathrm{eff;\,N}}^\mathrm{SNM}$           &   $m_{\mathrm{n}}$         & $0.624$ & $^{+0.013}_{-0.013}$ & $0.634$  & $^{+0.014}_{-0.015}$   & $0.629$  & $^{+0.012}_{-0.014}$   & $0.625$ & $0.629$ & $0.638$ \\
$m_{\mathrm{eff;\,n}}^\mathrm{PNM}$           &   $m_{\mathrm{n}}$         & $0.839$ & $^{+0.012}_{-0.014}$ & $0.852$  & $^{+0.017}_{-0.018}$   & $0.839$  & $^{+0.012}_{-0.013}$   & $0.831$ & $0.840$ & $0.839$ \\
$M_\mathrm{G}^*$                           &   $\Msun$                & $2.09$  & $^{+0.084}_{-0.038}$ & $2.08$   & $^{+0.071}_{-0.033}$   & $2.07$   & $^{+0.068}_{-0.036}$   & $2.04$ & $2.04$ & $2.08$ \\
$M_\mathrm{B}^*$                           &   $\Msun$                & $2.48$  & $^{+0.11}_{-0.049}$  & $2.5$    & $^{+0.087}_{-0.053}$   & $2.47$   & $^{+0.089}_{-0.046}$   & $2.43$ & $2.43$ & $2.47$ \\
$c^{*2}_\mathrm{s}$                        &   $c^2$	            & $0.883$  & $^{+0.068}_{-0.069}$ & $0.969$  & $^{+0.028}_{-0.077}$   & $0.955$  & $^{+0.035}_{-0.068}$   & $0.947$ & $0.905$ & $0.921$ \\
$n_\mathrm{c}^*$                           &   $\mathrm{fm}^{-3}$     & $1.13$   & $^{+0.04}_{-0.069}$  & $1.18$   & $^{+0.045}_{-0.079}$   & $1.16$   & $^{+0.043}_{-0.066}$   & $1.20$ & $1.17$ & $1.15$ \\
$\rho_\mathrm{c}^*$                        &   $10^{15}~\mathrm{g/cm^3}$  & $2.66$ & $^{+0.1}_{-0.16}$  & $2.8$    & $^{+0.12}_{-0.19}$    & $2.74$   & $^{+0.11}_{-0.16}$     & $2.83$ & $2.74$ & $2.70$ \\
$P_{\mathrm{c}}^*$                           &   $10^{36}~\mathrm{dyn/cm^2}$ & $1.2$ & $^{+0.061}_{-0.077}$ & $1.38$ & $^{+0.14}_{-0.14}$    & $1.27$  & $^{+0.071}_{-0.094}$    & $1.33$ & $1.23$ & $1.24$ \\
$R_{1.4}$                                  &   $\mathrm{km}$	    & $12.3$ & $^{+0.31}_{-0.22}$      & $11.8$  & $^{+0.47}_{-0.45}$    & $12.1$  & $^{+0.3}_{-0.21}$      & $12.0$ & $12.2$ & $12.2$ \\
$\Lambda_{1.4}$                            & 	--	            & $394$ & $^{+74}_{-43}$          & $306$   & $^{+90}_{-65}$        & $367$   & $^{+68}_{-38}$        & $303$  & $378$ & $379$ \\
$R_{2.0}$                                  &   $\mathrm{km}$	    & $11.3$ & $^{+0.55}_{-0.35}$      & $10.9$  & $^{+0.63}_{-0.37}$    & $11.1$  & $^{+0.53}_{-0.36}$     & $10.7$ & $10.9$ & $11.1$ \\
$\Lambda_{2.0}$                            & 	--	            & $18.3$ & $^{+9.6}_{-4.5}$        & $13.5$  & $^{+8.4}_{-3.4}$      & $15.8$  & $^{+8.2}_{-4.2}$      & $10.7$ & $12.4$ & $15.4$ \\
\bottomrule
\bottomrule
\end{tabular}
\end{table*}
\renewcommand{\arraystretch}{1.0}
\setlength{\tabcolsep}{2.0pt}
%%%%%%%%%%%%%%%%%%%%%%%%%%%%%%%%%%%%%%%%%%%%%%%%%%%%%%%%%%%%%%%%%%%%%%

%%%%%%%%%%%%%%%%%%%%%%%%%%%%%%%%%%%%%%%%%%%%%%%%%%%%%%%%%%%%%%%%%%%%%%
\bibliography{BSk_13.bib}

@article{Agrawal_PRC_2006,
    author = "Agrawal, B. K. and Dhiman, Shashi K. and Kumar, Raj",
    title = "{Exploring the extended density-dependent Skyrme effective forces for normal and isospin-rich nuclei to neutron stars}",
    eprint = "nucl-th/0604045",
    archivePrefix = "arXiv",
    doi = "10.1103/PhysRevC.73.034319",
    journal = "Phys. Rev. C",
    volume = "73",
    pages = "034319",
    year = "2006"
}

@article{Typel_EPJA_2022,
    author = "Typel, S. and others",
    collaboration = "CompOSE Core Team",
    title = "{CompOSE Reference Manual}",
    eprint = "2203.03209",
    archivePrefix = "arXiv",
    primaryClass = "astro-ph.HE",
    doi = "10.1140/epja/s10050-022-00847-y",
    journal = "Eur. Phys. J. A",
    volume = "58",
    number = "11",
    pages = "221",
    year = "2022"
}

@article{Beznogov_MNRAS_2015a,
    author = "Beznogov, M. V. and Yakovlev, D. G.",
    title = "{Statistical theory of thermal evolution of neutron stars}",
    eprint = "1411.6803",
    archivePrefix = "arXiv",
    primaryClass = "astro-ph.SR",
    doi = "10.1093/mnras/stu2506",
    journal = "Mon. Not. Roy. Astron. Soc.",
    volume = "447",
    number = "2",
    pages = "1598--1609",
    year = "2015"
}

@dataset{BSk_13Par_data,
	author       = {Beznogov, Mikhail V. and Raduta, Adriana R},
	title        = {Bayesian inference of the dense matter equation of state built upon extended Skyrme interactions: A generalization [Data Set]},
	month        = oct,
	year         = 2026,
	publisher    = {Zenodo},
	doi          = {10.5281/zenodo.23763922},
	url          = {https://doi.org/10.5281/zenodo.23763922},
}

@article{Beznogov_MNRAS_2015b,
    author = "Beznogov, M. V. and Yakovlev, D. G.",
    title = "{Statistical theory of thermal evolution of neutron stars {\textendash} II. Limitations on direct Urca threshold}",
    eprint = "1507.04206",
    archivePrefix = "arXiv",
    primaryClass = "astro-ph.SR",
    doi = "10.1093/mnras/stv1293",
    journal = "Mon. Not. Roy. Astron. Soc.",
    volume = "452",
    number = "1",
    pages = "540--548",
    year = "2015"
}

@article{Constantinou_PRC_2014,
    author = "Constantinou, Constantinos and Muccioli, Brian and Prakash, Madappa and Lattimer, James M.",
    title = "{Thermal properties of supernova matter: The bulk homogeneous phase}",
    eprint = "1402.6348",
    archivePrefix = "arXiv",
    primaryClass = "astro-ph.SR",
    doi = "10.1103/PhysRevC.89.065802",
    journal = "Phys. Rev. C",
    volume = "89",
    number = "6",
    pages = "065802",
    year = "2014"
}

@article{Constantinou_PRC_2015,
    author = "Constantinou, Constantinos and Muccioli, Brian and Prakash, Madappa and Lattimer, James M.",
    title = "{Thermal properties of hot and dense matter with finite range interactions}",
    eprint = "1504.03982",
    archivePrefix = "arXiv",
    primaryClass = "astro-ph.SR",
    doi = "10.1103/PhysRevC.92.025801",
    journal = "Phys. Rev. C",
    volume = "92",
    number = "2",
    pages = "025801",
    year = "2015"
}

@article{Fortin_PRC_2016,
    author = "Fortin, M. and Providencia, C. and Raduta, A. R. and Gulminelli, F. and Zdunik, J. L and Haensel, P. and Bejger, M.",
    title = "{Neutron star radii and crusts: uncertainties and unified equations of state}",
    eprint = "1604.01944",
    archivePrefix = "arXiv",
    primaryClass = "astro-ph.SR",
    doi = "10.1103/PhysRevC.94.035804",
    journal = "Phys. Rev. C",
    volume = "94",
    number = "3",
    pages = "035804",
    year = "2016"
}

@article{Drischler_PRC_2021,
  title = "{Constraints on the nuclear symmetry energy from asymmetric-matter calculations with chiral $NN$ and $3N$ interactions}",
  author = {Somasundaram, R. and Drischler, C. and Tews, I. and Margueron, J.},
  journal = {Phys. Rev. C},
  volume = {103},
  issue = {4},
  pages = {045803},
  numpages = {18},
  year = {2021},
  month = {Apr},
  publisher = {American Physical Society},
  doi = {10.1103/PhysRevC.103.045803},
  url = {https://link.aps.org/doi/10.1103/PhysRevC.103.045803}
}

@article{Margueron_PRC_2018a,
  title = "{Equation of state for dense nucleonic matter from metamodeling. I. Foundational aspects}",
  author = {Margueron, J\'er\^ome and Hoffmann Casali, Rudiney and Gulminelli, Francesca},
  journal = {Phys. Rev. C},
  volume = {97},
  issue = {2},
  pages = {025805},
  numpages = {28},
  year = {2018},
  month = {Feb},
  publisher = {American Physical Society},
  doi = {10.1103/PhysRevC.97.025805},
  url = {https://link.aps.org/doi/10.1103/PhysRevC.97.025805}
}

@article{Fonseca_2021,
	doi = {10.3847/2041-8213/ac03b8},
	url = {https://doi.org/10.3847/2041-8213/ac03b8},
	year = 2021,
	month = {jul},
	publisher = {American Astronomical Society},
	volume = {915},
	number = {1},
	pages = {L12},
	author = {E. Fonseca and others},
	title = "{Refined Mass and Geometric Measurements of the High-mass {PSR} J0740$+$6620}",
	journal = {Astrophys. J. Lett.},
}

@article{Beznogov_PRC_2024,
    author = "Beznogov, Mikhail V. and Raduta, Adriana R.",
    title = "{Bayesian inference of the dense matter equation~of state built upon extended Skyrme interactions}",
    eprint = "2403.19325",
    archivePrefix = "arXiv",
    primaryClass = "nucl-th",
    doi = "10.1103/PhysRevC.110.035805",
    journal = "Phys. Rev. C",
    volume = "110",
    number = "3",
    pages = "035805",
    year = "2024"
}

@article{Beznogov_ApJ_2024,
    author = "Beznogov, Mikhail V. and Raduta, Adriana R.",
    title = "{Bayesian Survey of the Dense Matter Equation of State Built upon Skyrme Effective Interactions}",
    eprint = "2308.15351",
    archivePrefix = "arXiv",
    primaryClass = "astro-ph.HE",
    doi = "10.3847/1538-4357/ad2f9b",
    journal = "Astrophys. J.",
    volume = "966",
    number = "2",
    pages = "216",
    year = "2024"
}

@article{Beznogov_PRC_2023,
    author = "Beznogov, Mikhail V. and Raduta, Adriana R.",
    title = "{Bayesian inference of the dense matter equation~of state built upon covariant density functionals}",
    eprint = "2212.07168",
    archivePrefix = "arXiv",
    primaryClass = "nucl-th",
    doi = "10.1103/PhysRevC.107.045803",
    journal = "Phys. Rev. C",
    volume = "107",
    number = "4",
    pages = "045803",
    year = "2023"
}

@article{Raduta_PLB_2024,
    author = "Raduta, Adriana R. and Beznogov, Mikhail V. and Oertel, Micaela",
    title = "{Bayesian inference of thermal effects in dense matter within the covariant density functional theory}",
    eprint = "2402.14593",
    archivePrefix = "arXiv",
    primaryClass = "nucl-th",
    doi = "10.1016/j.physletb.2024.138696",
    journal = "Phys. Lett. B",
    volume = "853",
    pages = "138696",
    year = "2024"
}

@article{Raduta_AA_2025,
    author = "Raduta, Adriana R. and Beznogov, Mikhail V.",
    title = "{New ab initio constrained extended Skyrme equations of state for simulations of neutron stars, supernovae, and binary mergers - I. Subsaturation density domain}",
    eprint = "2504.21725",
    archivePrefix = "arXiv",
    primaryClass = "nucl-th",
    doi = "10.1051/0004-6361/202555351",
    journal = "Astron. Astrophys.",
    volume = "701",
    pages = "A143",
    year = "2025"
}

@article{Raduta_AA_2026,
    author = "Raduta, Adriana R. and Beznogov, Mikhail V.",
    title = "{New ab initio constrained extended Skyrme equations of state for simulations of neutron stars, supernovae, and binary mergers - II. Thermal response in the suprasaturation density domain}",
    eprint = "2509.23910",
    archivePrefix = "arXiv",
    primaryClass = "nucl-th",
    doi = "10.1051/0004-6361/202557433",
    journal = "Astron. Astrophys.",
    volume = "705",
    pages = "A151",
    year = "2026"
}

@article{NV_1973,
	title = {Neutron star matter at sub-nuclear densities},
	journal = {Nucl. Phys. A},
	volume = {207},
	number = {2},
	pages = {298-320},
	year = {1973},
	issn = {0375-9474},
	doi = {https://doi.org/10.1016/0375-9474(73)90349-7},
	url = {https://www.sciencedirect.com/science/article/pii/0375947473903497},
	author = {J.W. Negele and D. Vautherin},
}

@ARTICLE{HDZ_1989,
       author = "{Haensel}, P. and {Zdunik}, J.~L. and {Dobaczewski}, J.",
        title = "{Composition and equation of state of cold catalyzed matter below neutron drip}",
      journal = {Astron. Astrophys.},
         year = 1989,
        month = sep,
       volume = {222},
       number = {1-2},
        pages = {353-357},
       adsurl = {https://ui.adsabs.harvard.edu/abs/1989A&A...222..353H}
}

@ARTICLE{Hartle_ApJ_1967,
       author = {{Hartle}, James B.},
        title = "{Slowly Rotating Relativistic Stars. I. Equations of Structure}",
      journal = {\apj},
         year = 1967,
        month = dec,
       volume = {150},
        pages = {1005},
          doi = {10.1086/149400},
       adsurl = {https://ui.adsabs.harvard.edu/abs/1967ApJ...150.1005H}
}

@article{Miller_ApJL_2021,
	doi = {10.3847/2041-8213/ac089b},
	url = {https://doi.org/10.3847/2041-8213/ac089b},
	year = 2021,
	month = {sep},
	publisher = {American Astronomical Society},
	volume = {918},
	number = {2},
	pages = {L28},
	author = {M. C. Miller and others},
	title = "{The Radius of {PSR} J0740+6620 from {NICER} and {XMM}-Newton Data}",
	journal = {Astrophys. J. Lett.},
   }

@ARTICLE{Abbott_PRL_2017,
       author = "Abbott, B. P. and others",
        title = "{GW170817: Observation of Gravitational Waves from a Binary Neutron Star Inspiral}",
      journal = {\prl},
         year = 2017,
        month = oct,
       volume = {119},
       number = {16},
          eid = {161101},
        pages = {161101},
          doi = {10.1103/PhysRevLett.119.161101},
archivePrefix = {arXiv},
       eprint = {1710.05832},
 primaryClass = {gr-qc},
       adsurl = {https://ui.adsabs.harvard.edu/abs/2017PhRvL.119p1101A}
}

@article{Abbott_PRL_121,
  title = {GW170817: Measurements of Neutron Star Radii and Equation of State},
  author = {{Abbott}, B.~P. and others},
  collaboration = {The LIGO Scientific Collaboration and the Virgo Collaboration},
  journal = {Phys. Rev. Lett.},
  volume = {121},
  issue = {16},
  pages = {161101},
  numpages = {16},
  year = {2018},
  month = {Oct},
  publisher = {American Physical Society},
  doi = {10.1103/PhysRevLett.121.161101},
  url = {https://link.aps.org/doi/10.1103/PhysRevLett.121.161101}
}

@article{Abbott_PRX_2019,
    author = "Abbott, B. P. and others",
    collaboration = "LIGO Scientific, Virgo",
    title = "{Properties of the binary neutron star merger GW170817}",
    eprint = "1805.11579",
    archivePrefix = "arXiv",
    primaryClass = "gr-qc",
    doi = "10.1103/PhysRevX.9.011001",
    journal = "Phys. Rev. X",
    volume = "9",
    number = "1",
    pages = "011001",
    year = "2019"
}

@article{Negele_PRC_1972,
  title = {Density-Matrix Expansion for an Effective Nuclear Hamiltonian},
  author = {Negele, J. W. and Vautherin, D.},
  journal = {Phys. Rev. C},
  volume = {5},
  issue = {5},
  pages = {1472--1493},
  numpages = {0},
  year = {1972},
  month = {May},
  publisher = {American Physical Society},
  doi = {10.1103/PhysRevC.5.1472},
  url = {https://link.aps.org/doi/10.1103/PhysRevC.5.1472}
}

@article{Chamel_PRC_2009,
  title = "{Further explorations of Skyrme--Hartree--Fock--Bogoliubov mass formulas. XI. Stabilizing neutron stars against a ferromagnetic collapse}",
  author = {Chamel, N. and Goriely, S. and Pearson, J. M.},
  journal = {Phys. Rev. C},
  volume = {80},
  issue = {6},
  pages = {065804},
  numpages = {12},
  year = {2009},
  month = {Dec},
  publisher = {American Physical Society},
  doi = {10.1103/PhysRevC.80.065804},
  url = {https://link.aps.org/doi/10.1103/PhysRevC.80.065804}
}

@article{Dutra_PRC_2012,
  title = {Skyrme interaction and nuclear matter constraints},
  author = {Dutra, M. and Louren\c{c}o, O. and S\'a Martins, J. S. and Delfino, A. and Stone, J. R. and Stevenson, P. D.},
  journal = {Phys. Rev. C},
  volume = {85},
  issue = {3},
  pages = {035201},
  numpages = {36},
  year = {2012},
  month = {Mar},
  publisher = {American Physical Society},
  doi = {https://doi.org/10.1103/PhysRevC.85.035201},
  url = {https://link.aps.org/doi/10.1103/PhysRevC.85.035201}
}

@article{Baldo_PRC_2014,
  title = "{Nucleon effective masses within the Brueckner-Hartree-Fock theory: Impact on stellar neutrino emission}",
  author = {Baldo, M. and Burgio, G. F. and Schulze, H.-J. and Taranto, G.},
  journal = {Phys. Rev. C},
  volume = {89},
  issue = {4},
  pages = {048801},
  numpages = {5},
  year = {2014},
  month = {Apr},
  publisher = {American Physical Society},
  doi = {10.1103/PhysRevC.89.048801},
  url = {https://link.aps.org/doi/10.1103/PhysRevC.89.048801}
}

@article{Burgio_PRC_2020,
  title = {Nucleon effective mass in hot dense matter},
  author = {Shang, X. L. and Li, A. and Miao, Z. Q. and Burgio, G. F. and Schulze, H.-J.},
  journal = {Phys. Rev. C},
  volume = {101},
  issue = {6},
  pages = {065801},
  numpages = {9},
  year = {2020},
  month = {Jun},
  publisher = {American Physical Society},
  doi = {10.1103/PhysRevC.101.065801},
  url = {https://link.aps.org/doi/10.1103/PhysRevC.101.065801}
}

@article{Urban_PRC_2023,
  title = "{Energy and angle dependence of neutrino scattering rates in proto--neutron star and supernova matter within Skyrme RPA}",
  author = {Duan, Mingya and Urban, Michael},
  journal = {Phys. Rev. C},
  volume = {108},
  issue = {2},
  pages = {025813},
  numpages = {16},
  year = {2023},
  month = {Aug},
  publisher = {American Physical Society},
  doi = {10.1103/PhysRevC.108.025813},
  url = {https://link.aps.org/doi/10.1103/PhysRevC.108.025813}
}

@ARTICLE{Antoniadis_Science_2013,
     author = "Antoniadis, John and others",
        title = "{A Massive Pulsar in a Compact Relativistic Binary}",
      journal = {Science},
         year = 2013,
        month = apr,
       volume = {340},
       number = {6131},
        pages = {448},
          doi = {10.1126/science.1233232},
archivePrefix = {arXiv},
       eprint = {1304.6875},
 primaryClass = {astro-ph.HE},
       adsurl = {https://ui.adsabs.harvard.edu/abs/2013Sci...340..448A}
}

@Article{Cromartie_Nature_2020,
  author = "Cromartie, H. T. and others",
 title={Relativistic Shapiro delay measurements of an extremely massive millisecond pulsar},
journal={Nature Astronomy},
year={2020},
month={Jan},
day={01},
volume={4},
number={1},
pages={72-76},
issn={2397-3366},
doi={10.1038/s41550-019-0880-2},
url={https://doi.org/10.1038/s41550-019-0880-2}
}

@article{Riley_ApJL_2019,
doi = {10.3847/2041-8213/ab481c},
url = {https://doi.org/10.3847/2041-8213/ab481c},
year = {2019},
month = {dec},
publisher = {The American Astronomical Society},
volume = {887},
number = {1},
pages = {L21},
author = "Riley, T. E. and others",
title = {A NICER View of PSR J0030+0451: Millisecond Pulsar Parameter Estimation},
journal = {The Astrophysical Journal Letters},
}

@article{Riley_ApJL_2021,
doi = {10.3847/2041-8213/ac0a81},
url = {https://doi.org/10.3847/2041-8213/ac0a81},
year = {2021},
month = {sep},
publisher = {The American Astronomical Society},
volume = {918},
number = {2},
pages = {L27},
author = "Riley, T. E. and others",
title = {A NICER View of the Massive Pulsar PSR J0740+6620 Informed by Radio Timing and XMM-Newton Spectroscopy},
journal = {The Astrophysical Journal Letters},
}

@article{Miller_ApJ_2019,
      author         = "Miller, M. C. and others",
      title          = "{PSR J0030+0451 Mass and Radius from NICER Data and
                        Implications for the Properties of Neutron Star Matter}",
      journal        = "Astrophys. J. Lett.",
      volume         = "887",
      year           = "2019",
      pages          = "L24",
      doi            = "10.3847/2041-8213/ab50c5",
       SLACcitation   = "%%CITATION = ARXIV:1912.05705;%%"
}

@article{Salmi_ApJ_2022,
    author = "Salmi, Tuomo and others",
    title = "{The Radius of PSR J0740+6620 from NICER with NICER Background Estimates}",
    eprint = "2209.12840",
    archivePrefix = "arXiv",
    primaryClass = "astro-ph.HE",
    doi = "10.3847/1538-4357/ac983d",
    journal = "Astrophys. J.",
    volume = "941",
    number = "2",
    pages = "150",
    year = "2022"
}

@article{Salmi_ApJ_2024b,
    author = "Salmi, Tuomo and others",
    title = "{A NICER View of PSR J1231{\ensuremath{-}}1411: A Complex Case}",
    eprint = "2409.14923",
    archivePrefix = "arXiv",
    primaryClass = "astro-ph.HE",
    doi = "10.3847/1538-4357/ad81d2",
    journal = "Astrophys. J.",
    volume = "976",
    number = "1",
    pages = "58",
    year = "2024"
}

@article{Miller_ApJ_2026,
    author = "Miller, M. C. and others",
    title = "{The Radius of PSR J0437{\textendash}4715 from NICER Data}",
    eprint = "2512.08790",
    archivePrefix = "arXiv",
    primaryClass = "astro-ph.HE",
    doi = "10.3847/2041-8213/ae5057",
    journal = "Astrophys. J. Lett.",
    volume = "1000",
    number = "2",
    pages = "L48",
    year = "2026"
}

@article{Vinciguerra_ApJ_2024,
    author = "Vinciguerra, Serena and others",
    title = "{An Updated Mass{\textendash}Radius Analysis of the 2017{\textendash}2018 NICER Data Set of PSR J0030+0451}",
    eprint = "2308.09469",
    archivePrefix = "arXiv",
    primaryClass = "astro-ph.HE",
    doi = "10.3847/1538-4357/acfb83",
    journal = "Astrophys. J.",
    volume = "961",
    number = "1",
    pages = "62",
    year = "2024"
}

@article{Papakonstantinou_PRC_2023,
    author = "Zhou, Jia and Xu, Jun and Papakonstantinou, Panagiota",
    title = "{Bayesian inference of neutron-star observables based on effective nuclear interactions}",
    eprint = "2301.07904",
    archivePrefix = "arXiv",
    primaryClass = "nucl-th",
    doi = "10.1103/PhysRevC.107.055803",
    journal = "Phys. Rev. C",
    volume = "107",
    number = "5",
    pages = "055803",
    year = "2023"
}

@ARTICLE{Imam_PRD_2024,
       author = {{Imam}, Sk Md Adil and {Malik}, Tuhin and {Provid{\^e}ncia}, Constan{\c{c}}a and {Agrawal}, B.~K.},
        title = "{Implications of comprehensive nuclear and astrophysics data on the equations of state of neutron star matter}",
      journal = {Phys. Rev. D},
         year = 2024,
        month = may,
       volume = {109},
       number = {10},
          eid = {103025},
        pages = {103025},
          doi = {10.1103/PhysRevD.109.103025},
archivePrefix = {arXiv},
       eprint = {2401.06018},
 primaryClass = {nucl-th},
       adsurl = {https://ui.adsabs.harvard.edu/abs/2024PhRvD.109j3025I}
}

@article{Traversi_ApJ_2020,
doi = {10.3847/1538-4357/ab99c1},
url = {https://dx.doi.org/10.3847/1538-4357/ab99c1},
year = {2020},
month = {jul},
publisher = {The American Astronomical Society},
volume = {897},
number = {2},
pages = {165},
author = {Silvia Traversi and Prasanta Char and Giuseppe Pagliara},
title = {Bayesian Inference of Dense Matter Equation of State within Relativistic Mean Field Models Using Astrophysical Measurements},
journal = {Astrophys. J.},
}

@article{Malik_ApJ_2022,
    author = "Malik, Tuhin and Ferreira, M\'arcio and Agrawal, B. K. and Provid\^encia, Constan\c{c}a",
    title = "{Relativistic Description of Dense Matter Equation of State and Compatibility with Neutron Star Observables: A Bayesian Approach}",
     doi = "10.3847/1538-4357/ac5d3c",
    journal = "Astrophys. J.",
    volume = "930",
    number = "1",
    pages = "17",
    year = "2022"
}

@article{Malik_PRD_2023,
  title = {Spanning the full range of neutron star properties within a microscopic description},
  author = {Malik, Tuhin and Ferreira, M\'arcio and Albino, Milena Bastos and Provid\^encia, Constan\c{c}a},
  journal = {Phys. Rev. D},
  volume = {107},
  issue = {10},
  pages = {103018},
  numpages = {18},
  year = {2023},
  month = {May},
  publisher = {American Physical Society},
  doi = {10.1103/PhysRevD.107.103018},
  url = {https://link.aps.org/doi/10.1103/PhysRevD.107.103018}
}

@article{Char_PRD_2023,
  title = {Generalized description of neutron star matter with a nucleonic relativistic density functional},
  author = {Char, P. and Mondal, C. and Gulminelli, F. and Oertel, M.},
  journal = {Phys. Rev. D},
  volume = {108},
  issue = {10},
  pages = {103045},
  numpages = {15},
  year = {2023},
  month = {Nov},
  publisher = {American Physical Society},
  doi = {10.1103/PhysRevD.108.103045},
  url = {https://link.aps.org/doi/10.1103/PhysRevD.108.103045}
}

@article{Cartaxo_ApJSS_2026,
    author = "Cartaxo, Jo{\~a}o and Huang, Chun and Malik, Tuhin and Sourav, Shashwat and Yuan, Wen-Li and Zhou, Tianzhe and Liu, Xuezhi and Provid{\^e}ncia, Constan{\c{c}}a",
    title = "{Covariant Energy Density Functionals for Modeling the Equation of State of Neutron Star Matter: Cross-comparison Analysis Using CompactObject}",
    eprint = "2506.03112",
    archivePrefix = "arXiv",
    primaryClass = "nucl-th",
    reportNumber = "ET-0331A-25",
    doi = "10.3847/1538-4365/ae2310",
    journal = "Astrophys. J. Suppl.",
    volume = "282",
    number = "2",
    pages = "33",
    year = "2026"
}

@article{Li_PRC_2025,
    author = "Li, Jia-Jie and Tian, Yu and Sedrakian, Armen",
    title = "{Bayesian inferences on covariant density functionals from multimessenger astrophysical data: Nucleonic models}",
    eprint = "2502.20000",
    archivePrefix = "arXiv",
    primaryClass = "nucl-th",
    doi = "10.1103/PhysRevC.111.055804",
    journal = "Phys. Rev. C",
    volume = "111",
    number = "5",
    pages = "055804",
    year = "2025"
}

@ARTICLE{BSk19-BSk21,
       author = {{Goriely}, S. and {Chamel}, N. and {Pearson}, J.~M.},
        title = "{Further explorations of Skyrme-Hartree-Fock-Bogoliubov mass formulas. XII. Stiffness and stability of neutron-star matter}",
      journal = {Phys. Rev. C},
         year = 2010,
        month = sep,
       volume = {82},
       number = {3},
          eid = {035804},
        pages = {035804},
          doi = {10.1103/PhysRevC.82.035804},
archivePrefix = {arXiv},
       eprint = {1009.3840},
 primaryClass = {nucl-th},
       adsurl = {https://ui.adsabs.harvard.edu/abs/2010PhRvC..82c5804G}
}

@ARTICLE{BSk22-BSk26,
       author = {{Goriely}, S. and {Chamel}, N. and {Pearson}, J.~M.},
        title = "{Further explorations of Skyrme-Hartree-Fock-Bogoliubov mass formulas. XIII. The 2012 atomic mass evaluation and the symmetry coefficient}",
      journal = {Phys. Rev. C},
         year = 2013,
        month = aug,
       volume = {88},
       number = {2},
          eid = {024308},
        pages = {024308},
          doi = {10.1103/PhysRevC.88.024308},
       adsurl = {https://ui.adsabs.harvard.edu/abs/2013PhRvC..88b4308G}
}

@ARTICLE{BSk29-BSk32,
       author = {{Goriely}, S. and {Chamel}, N. and {Pearson}, J.~M.},
        title = "{Further explorations of Skyrme-Hartree-Fock-Bogoliubov mass formulas. XVI. Inclusion of self-energy effects in pairing}",
      journal = {Phys. Rev. C},
         year = 2016,
        month = mar,
       volume = {93},
       number = {3},
          eid = {034337},
        pages = {034337},
          doi = {10.1103/PhysRevC.93.034337},
       adsurl = {https://ui.adsabs.harvard.edu/abs/2016PhRvC..93c4337G}
}

@ARTICLE{emcee,
        author = {{Foreman-Mackey}, Daniel and {Hogg}, David W. and {Lang}, Dustin and {Goodman}, Jonathan},
        title = "{emcee: The MCMC Hammer}",
        journal = {Publ. Astron. Soc. Pac.},
        keywords = {Astrophysics - Instrumentation and Methods for Astrophysics, Physics - Computational Physics, Statistics - Computation},
        year = {2013},
        month = {mar},
        volume = {125},
        number = {925},
        pages = {306},
        doi = {10.1086/670067},
        !archivePrefix = {arXiv},
        !eprint = {1202.3665},
        !primaryClass = {astro-ph.IM},
        adsurl = {https://ui.adsabs.harvard.edu/abs/2013PASP..125..306F},
        adsnote = {Provided by the SAO/NASA Astrophysics Data System}
}

@ARTICLE{ptemcee,
	author = {{Vousden}, W.~D. and {Farr}, W.~M. and {Mandel}, I.},
	title = "{Dynamic temperature selection for parallel tempering in Markov chain Monte Carlo simulations}",
	journal = {Mon. Not. R. Astron. Soc.},
	keywords = {methods: data analysis, methods: numerical, methods: statistical, Astrophysics - Instrumentation and Methods for Astrophysics},
	year = 2016,
	month = jan,
	volume = {455},
	number = {2},
	pages = {1919-1937},
	doi = {10.1093/mnras/stv2422},
	!archivePrefix = {arXiv},
	!eprint = {1501.05823},
	!primaryClass = {astro-ph.IM},
	adsurl = {https://ui.adsabs.harvard.edu/abs/2016MNRAS.455.1919V},
	adsnote = {Provided by the SAO/NASA Astrophysics Data System}
}

@INPROCEEDINGS{Skilling_2004,
        author = {{Skilling}, John},
        title = "{Nested Sampling}",
        booktitle = {Bayesian Inference and Maximum Entropy Methods in Science and Engineering: 24th International Workshop on Bayesian Inference and Maximum Entropy Methods in Science and Engineering},
        year = 2004,
        editor = {{Fischer}, Rainer and {Preuss}, Roland and {Toussaint}, Udo Von},
        series = {American Institute of Physics Conference Series},
        volume = {735},
        month = nov,
        pages = {395-405},   
        doi = {10.1063/1.1835238},
        adsurl = {https://ui.adsabs.harvard.edu/abs/2004AIPC..735..395S}
}

@article{Skilling_2006,
        author = {{Skilling}, John},
        title = {{Nested sampling for general Bayesian computation}},
        volume = {1},
        journal = {Bayesian Analysis},
        number = {4},
        publisher = {International Society for Bayesian Analysis},
        pages = {833 -- 859},
        year = {2006},
        doi = {10.1214/06-BA127},
        URL = {https://doi.org/10.1214/06-BA127}
}

@ARTICLE{Buchner_2016,
       author = {{Buchner}, Johannes},
        title = "{A statistical test for Nested Sampling algorithms}",
      journal = {Stat.  Comput.},
         year = 2016,
        month = jan,
       volume = {26},
       number = {1-2},
        pages = {383-392},
          doi = {10.1007/s11222-014-9512-y},
archivePrefix = {arXiv},
      eprint = {1407.5459},
primaryClass = {stat.CO},
       adsurl = {https://ui.adsabs.harvard.edu/abs/2016S&C....26..383B}
}

@ARTICLE{Buchner_2019,
       author = {{Buchner}, Johannes},
        title = "{Collaborative Nested Sampling: Big Data versus Complex Physical Models}",
      journal = {Publ. Astron. Soc. Pac.},
         year = 2019,
        month = oct,
       volume = {131},
       number = {1004},
        pages = {108005},
          doi = {10.1088/1538-3873/aae7fc},
archivePrefix = {arXiv},
      eprint = {1707.04476},
primaryClass = {stat.CO},
       adsurl = {https://ui.adsabs.harvard.edu/abs/2019PASP..131j8005B}
}

@ARTICLE{ultranest,
       author = {{Buchner}, Johannes},
        title = "{UltraNest - a robust, general purpose Bayesian inference engine}",
      journal = {J. Open Source Software},
         year = 2021,
        month = apr,
       volume = {6},
       number = {60},
          eid = {3001},
        pages = {3001},
          doi = {10.21105/joss.03001},
archivePrefix = {arXiv},
      eprint = {2101.09604},
primaryClass = {stat.CO},
       adsurl = {https://ui.adsabs.harvard.edu/abs/2021JOSS....6.3001B}
}
%%%%%%%%%%%%%%%%%%%%%%%%%%%%%%%%%%%%%%%%%%%%%%%%%%%%%%%%%%%%%%%%%%%%%%
\end{document}